\documentclass[10pt]{iopart}
\usepackage{epsfig}
\usepackage{amssymb}
\usepackage{graphics}
\usepackage{graphicx}
\usepackage{psfrag}
\usepackage{showidx}
\makeindex

\begin{document}

\newtheorem{theorem}{Theorem}[section]
\newtheorem{lemma}[theorem]{Lemma}
\newtheorem{proposition}[theorem]{Proposition}
\newtheorem{corollary}[theorem]{Corollary}

\newenvironment{proof}[1][Proof:]{\begin{trivlist}
\item[\hskip \labelsep {\bfseries #1}]}{\end{trivlist}}

\newcommand{\BEQ}{\begin{equation}}     
\newcommand{\BEA}{\begin{eqnarray}}
\newcommand{\EEQ}{\end{equation}}       
\newcommand{\EEA}{\end{eqnarray}}
\newcommand{\nn}{\nonumber}
\newcommand{\bb}{\begin{eqnarray}}
\newcommand{\ee}{\end{eqnarray}}
\newcommand{\Vol}{{\cal A}}
\newcommand{\sign}{{\rm sign}}

\newcommand{\dif}{\Gamma}
\newcommand{\pr}{{\rm Pr}}
\newcommand{\binom}[2]{{#1 \choose #2}}
\newcommand{\coef}{{\cal S}}
\newcommand{\diffab}[2]{\partial_{#2} #1}
\newcommand{\diffabc}[3]{\partial_{#2#3} #1}
\newcommand{\Lo}{{\ell_0}}
\newcommand{\erfc}{{\rm erfc}}
\newcommand{\erf}{{\rm erf}}
\newcommand{\lar}{\stackrel{\curvearrowleft}{ }}
\newcommand{\rar}{\stackrel{\curvearrowright}{ }}
\newcommand{\conc}{c}
\newcommand{\we}[1]{{\cal W}_{\ell_0}{(\mbox{\small $#1$})}}
\newcommand{\wet}[1]{\tilde\we{#1}}
\newcommand{\wesl}[1]{{\cal W}{(\mbox{\small $#1$})}}
\newcommand{\wetsl}[1]{\tilde\we{#1}}
\newcommand{\eps}{\varepsilon}          
\newcommand{\vph}{\varphi}              
\newcommand{\vth}{\vartheta}            
\newcommand{\D}{{\rm d}}
\newcommand{\II}{{\rm i}}               
\newcommand{\arcosh}{{\rm arcosh\,}}    
\newcommand{\wit}[1]{\widetilde{#1}}    
\newcommand{\wht}[1]{\widehat{#1}}      
\newcommand{\lap}[1]{\overline{#1}}     

\newcommand{\appsektion}[1]{\setcounter{equation}{0} \section*{Appendix. #1}
\renewcommand{\theequation}{A\arabic{equation}}
              \renewcommand{\thesection}{A} }

\newcommand{\appsection}[2]{\setcounter{equation}{0} \section*{Appendix #1. #2}
\renewcommand{\theequation}{#1\arabic{equation}}
              \renewcommand{\thesection}{#1} }

\title[Coagulation-diffusion process in $1D$ by the empty-interval method]{\textbf{Exact correlations 
in the one-dimensional coa\-gu\-la\-tion-dif\-fu\-sion process \\ by the empty-interval method}}

\author{Xavier Durang$^a$, Jean-Yves Fortin$^a$, Diego Del Biondo\footnote{adresse actuelle: 
Laboratoire de Physique de la Mati\`ere Condens\'ee et Nanostructures,
Universit\'e Claude Bernard Lyon 1 and CNRS, Domaine Scientifique de la Doua, B\^atiment L\'eon Brillouin,
43 Boulevard du 11 Novembre 1918, F - 69622 Villeurbanne, France}$^{a}$, \\ 
Malte Henkel$^a$ and Jean Richert$^{b}$}

\address{$^a$Groupe de Physique Statistique,
D\'epartement de Physique de la Mati\`ere et des Mat\'eriaux, 
Institut Jean Lamour\footnote{Laboratoire associ\'e au CNRS UMR 7198},
CNRS -- Nancy-Universit\'e -- UPVM, \\
B.P. 70239, F - 54506 Vand{\oe}uvre les Nancy Cedex, France\\
$^b$Institut de Physique, Universit\'e de Strasbourg, 3 rue de l'Universit\'e, F - 67084 Strasbourg Cedex
}
\ead{durang@lpm.u-nancy.fr,fortin@lpm.u-nancy.fr,delbiond@lpmcn.univ-lyon1.fr,\\
henkel@lpm.u-nancy.fr,richert@fresnel.u-strasbg.fr}

\date{\today}

\begin{abstract}
The long-time dynamics of reaction-diffusion processes in low dimensions is dominated by fluctuation effects.
The one-dimensional coagulation-diffusion process describes the kinetics of particles which freely hop between the
sites of a chain and where upon encounter of two particles, one of them disappears with probability one.
The empty-interval method has, since a long time, been a convenient tool
for the exact calculation of time-dependent particle densities in this model.
We generalize the empty-interval method by considering the probability distributions of two simultaneous empty
intervals at a given distance. While the equations of motion of these probabilities reduce for the
coagulation-diffusion process to a simple diffusion equation in the continuum limit,
consistency with the single-interval distribution
introduces several non-trivial boundary conditions which are solved for the first time for arbitrary initial 
configurations. In this way, exact space-time-dependent correlation functions can be directly obtained and
their dynamic scaling behaviour is analysed for large classes of initial conditions.
\end{abstract}

\pacs{05.20-y, 64.60.Ht}

~\\ {\bf Keywords:} exact results, phase transitions into absorbing states, diffusion, correlation functions  

\maketitle

\date{\today}

\section{Introduction}

The precise description of cooperative effects in strongly interacting many-body systems continues
to pose many challenges. Paradigmatic examples are systems which may be described in terms of
diffusion-limited reaction-diffusion processes. Applications of these
systems and their non-equilibrium phase transitions
have arisen in fields as different as solid-state physics, physical chemistry,
physical and chemical ageing, cosmology, biology, financial markets or population evolution in social sciences.
If the spatial dimension of these systems is low enough, that is $d\leq d^*$ where $d^*$ is the upper critical
dimension,  fluctuation effects dominate the long-time kinetics of these systems and their behaviour is different from
the one expected from the solutions of (mean-field) reaction-diffusion equations, which attempt to describe
the interactions of the elementary constituents in terms of the macroscopic law of mass-action, see e.g.
\cite{priv,Marr99,sch,benA00,Odor08,Henk09a}.

One of the motivations for this work is the continuing practical interest in systems with reduced dimensionality, and
such that homogenisation through stirring is not possible.
Specifically, we shall consider the one-dimensional {\em coagulation-diffusion process}, which is defined as follows.
Consider a single species $A$ of indistinguishable particles, such that each site of an infinitely long chain can
either be empty or else be occupied by a single particle. The dynamics of the system is described in terms of
a Markov process, where allowed two-site microscopic reactions $A+\emptyset \leftrightarrow \emptyset +A$ and
$A+A\to A+\emptyset$ or $\emptyset+A$ are implemented as follows: at each microscopic time step,
a randomly selected single particle hops to a
nearest-neighbour site, with a rate $\Gamma :=D a^2$,
where $a$ is the lattice constant. If that site was empty, the particle
is placed there. On the other hand, if the site was already occupied, one of the two particles is removed from the
system with probability one. This model is one of the best-studied examples of a diffusion-limited process and at
least since the work of Toussaint and Wilczek \cite{Tous83} it is known that
the mean particle concentration $c(t)\sim t^{-1/2}$ for large times and with an amplitude which is thought to be
universal as confirmed by the field-theoretical renormalisation group \cite{Lee94,Balb95};
in contrast a mean-field treatment would have
predicted $c(t)\sim t^{-1}$. These theoretically predicted fluctuation effects have been confirmed
experimentally, for example using the kinetics of excitons on long chains of the polymer
TMMC = (CH$_3$)$_4$N(MnCl$_3$) \cite{kro}, but also in other polymers confined to quasi-one-dimensional geometries
\cite{Pras89,Kope90}, see also the reviews in \cite{priv}. Another recent application of diffusion-limited reactions
concerns carbon nanotubes,
for example the relaxation of photoexcitations \cite{Russ06} or the photoluminescence
saturation \cite{Sriv09}. On the other hand, the $1D$ coagulation-diffusion process
has also received attention from mathematicians \cite{Bram80,Muna06a} 
and is simple enough that it can be related to integrable quantum chains, see \cite{Alca94,sch}. Hence, by
a consideration of the quantum chain Hamiltonians, which can be derived from the master equation, the
time-dependence of its observables could in principle be found via a Bethe ansatz, see \cite{Rosh05}.
In practice, however, it has turned
out to be easier to find the time-dependent densities from the {\em empty-interval method}, which
considers the time-dependent probabilities $E_n(t)$ that $n\geq 1$ consecutive sites of the chain are empty
\cite{benA90,Doer92,benA98,benA00,mas}, see also \cite{Spouge88}.
The $E_n(t)$ satisfy a closed set of differential-difference equations, subject to
the boundary condition $E_0(t)=1$ and the average particle concentration is obtained as
$c(t)=\bigl(1-E_1(t)\bigr)/a$. The scaling behaviour of the averages can be directly studied in the continuum limit
$a\to 0$, when $E_n(t)\longrightarrow E(x,t)$ which in turn satisfies the diffusion equation
$\bigl(\partial_t - 2D\partial_x^2\bigr)E(x,t)=0$ with the boundary
condition $E(0,t)=1$ such that the concentration now becomes $c(t)=\bigl.-\partial_x E(x,t)\bigr|_{x=0}$.
Still, the direct
solution of the problem is usually considered to be complicated enough to prefer to consider
instead $\rho(x,t) := \partial_x^2 E(x,t)$ where the boundary condition becomes $\rho(0,t)=0$ such that
standard Green's functions of the diffusion equation can be used, see \cite{benA00} and references therein.

Remarkably, the empty-interval method can be applied to a large class of coagulation-diffusion models, where several
additional reactions can be added, see e.g. 
\cite{benA90,Doer92,benA98,benA00,mas,Mobi01,Henk01,Khor03,Agha05,Muna06a,Muna06b,Maye07}. Furthermore, the quantum 
hamiltonian/Liouvillian of coagulation-diffusion models with the (reversible) reaction $2A\leftrightarrow A$ can, by a
stochastic similarity transformation \cite{Kreb95,Henk95,Dahm95}, be transformed to the one of pair annihilation/creation 
$2A\leftrightarrow \emptyset$ which in one dimension can be solved by free-fermion methods, see e.g. 
\cite{Bram80,Torn83,Racz85,Lush86,Alca94,Gryn95,Spouge88,Mobi02}. Those one-dimensional reaction-diffusion 
systems which can be treated with free-fermion methods have been classified \cite{Henk97,sch}, but the empty-interval 
technique has the advantage that further reactions can be treated, such as $\emptyset\longrightarrow A$ or 
$A\emptyset A\longrightarrow AAA$, which have no known analogue in a free-fermion description. 
In particular,
Peschel {\it et al.} \cite{pesch} suggested a systematic way to identify observables for which closed
systems of equations of motion can be derived from the reformulation of the master equation in terms of a
Hamiltonian matrix in a controllable way. Their approach includes the method of empty intervals as the
most simple special case. In principle, their method can be extended to include the probabilities of having
several empty intervals of sizes $n_1, n_2,\ldots$ at certain distances which allows to find correlation functions as
well. Their study is the main subject of this paper. 

In particular, our approach allows to consider {\em arbitrary} initial
configurations of particules and hence our results will include many of the existing results in the literature 
as special cases. As we shall see, there exists a natural decomposition of the time-dependent observables which
may be arranged in terms of the information required on the initial state. This can be formulated through 
single-interval or two-interval probabilities for those quantities which we consider explicitly. We shall give
examples which suggest a clear order to relevance in the long-time limit. On the other hand, we shall assume spatial 
translation-invariance from the outset, which simplifies
the equations to be analysed. However, if one were to investigate the effects of disorder, one would have to 
revert to a formalism \cite{Doer92,benA98} where translation-invariance is not required. 

The study of correlation functions of reaction-diffusion systems is also motivated by the recent interest in
ageing phenomena: having begun in the study of slow relaxation in glassy systems brought out of equilibrium after
a rapid change in the thermodynamic parameters, it was later realised that the three main characteristics of physical
ageing, namely (i) slow, non-exponential relaxation, (ii) breaking of time-translation-invariance and (iii)
dynamical scaling also occur in many-body systems which in contrast to glasses are neither disordered nor frustrated,
see \cite{Godr02} for a brief review and a forthcoming book \cite{Henk10}. 
Furthermore, these characteristics have also been found in several many-particle systems with {\em absorbing}
stationary states, such as the contact process \cite{Enss04,Rama04,Baum07}, the non-equilibrium kinetic Ising model
\cite{Odor06} or kinetically constrained systems such as the Frederikson-Andersen model \cite{Maye07}. 
One particular point of interest in these ageing systems is the relation between two-time correlations
and responses and a study of the coagulation-diffusion process (along with its exactly solved extensions)
should be useful, since
exact results can be expected, at least in one dimension. However, while such an analysis is readily formulated in
terms of the empty-interval method at two different times, the explicit calculation requires the knowledge of the
exact equal-time two-interval probabilities. In this paper, we shall provide this information, which will become an 
initial condition for the two-time correlator, and is going to be used
in a sequel paper where the ageing behaviour in exactly solvable reaction-diffusion processes will be addressed.

This paper is organised as follows. In order to make the presentation more self-contained,
we recall in section 2 the derivation of the equation of motion for the empty single-interval probability $E_n(t)$
before we proceed to show that the boundary condition $E_0(t)=1$ can be fixed through an
analytic continuation to negative values of $n$. The techniques thereby developed are to be generalized to the
two-interval probability in the remainder of the paper. The passage from the initial state towards the scaling long-time 
regime as a function of the initial distrbution is analysed and we also compare between the discrete model and its
continuum limit. In section~3, the equations of motion and the formal
solution is given, to be followed by the derivation of the consistency conditions with the single-interval
probabilities. The general two-interval probability for arbitrary initial conditions is derived in section~4 and
in section~5 we use the results for the derivation of the equal-time correlators. We conclude in section~6.
Several appendices (A-G) contain technical details of the calculations. 

\section{Single-interval probability}

\subsection{Equations of motion}

Using the definition of the coagulation-diffusion process as given in the introduction, we begin by recalling the
derivation of the equation of the empty-interval probabilities \cite{benA90}. The same equations can also be found
within a quantum Hamiltonian formalism \cite{pesch}, but this will not be repeated here.
We denote by $E_n(t)$ the time-dependent probability of having an interval of $n$
consecutive empty sites at time $t$.  Since the system is assumed to be homogeneous, $E_n(t)$
is site-independent and will depend only on the interval size $n$ and time $t$. The time
evolution of this quantity is governed by the rate at which particles move on adjacent
intervals of size $n$ or $n-1$.
In an interval of length $n$, which will be denoted by $\fbox{ n }$, a particle ($\bullet$)
can enter from the left or the right between the time period $t$ and $t+\D t$, and $E_n(t)$  decreases
during this period of time  by the amount
\bb\nn
-\Big [
\pr(\bullet\rar\fbox{ n })+\pr(\fbox{ n }\lar\bullet)
\Big ]
\ee
The probability $\pr(\bullet\rar\fbox{ n })$ is proportional to the
probability that a particle lies on the left of the interval, or
$\pr(\bullet\rar\fbox{ n })=\pr(\bullet\,\fbox{ n }) \Gamma \D t$,
which can be evaluated using the relation
\bb\nn
\pr(\bullet\,\fbox{ n })+\pr(\circ\,\fbox{ n }
)=\pr(\fbox{ n })=E_n(t)
\ee
where the symbol ($\circ$) refers to an empty site.
Since by definition $\pr(\circ\,\fbox{ n })=E_{n+1}(t)$ we obtain directly
\bb
\pr(\bullet\,\fbox{ n })=\pr(\fbox{ n }\,\bullet)=E_{n}(t)-E_{n+1}(t).
\ee
$E_n(t)$ may also increase, if we consider the possibility that a particle sitting next to an
interval of size $n-1$ moves away from this interval,
\bb\nn
+\Big [\pr(\lar \bullet\fbox{n-1})
+\pr( \fbox{n-1}\bullet\rar )
\Big ].
\ee
This is possible because the process $A+A\rightarrow A$ constrains each site
to contain at most one particle. Hence there is no need to
consider the case when the particle encounters another particle when it moves
away from the interval.
As before, we have
$\pr(\lar \bullet\fbox{n-1})=\pr(\bullet\,\fbox{n-1})\Gamma\D t = \bigl( E_{n-1}(t)-E_n(t)\bigr) \Gamma \D t$.
Summing the contributions, the rate of change for $E_n(t)$ is given by 
\bb\nn
\diffab{E_n(t)}{t}&=&2\Gamma
\Big [
-\{
E_{n}(t)-E_{n+1}(t) \}+
\{E_{n-1}(t)-E_{n}(t)\}
\Big ]
\\  \label{E1disc}
&=&
2\Gamma
\left (
E_{n-1}-2E_n+E_{n+1} \right ).
\ee
This equation is valid only for a positive index $n>1$. For $n=1$ the
rate of change for $E_1(t)$ is given as previously by the equation
\bb\nn
\diffab{E_1(t)}{t}\D t=\Big [
\pr (\bullet\rar)+\pr (\lar\bullet)-\pr(\bullet\rar\circ)
-\pr(\circ\lar\bullet)
\Big ].
\ee
We also have $\pr(\bullet\rar)=\pr(\bullet)\Gamma \D t$ and
$\pr(\bullet\rar\circ)=\pr(\bullet\,\circ)\Gamma \D t$. The solutions for each
of these quantities can be found by considering the probability conditions
\bb\nn
\pr(\bullet)&+&\pr(\circ)=1\;\Rightarrow\pr(\bullet)=1-E_1(t)
\\ \label{E1}
\pr(\bullet\,\circ)&+&\pr(\circ\,\circ)=\pr(\circ)\;\Rightarrow\pr(\bullet\,
\circ)=E_1(t)-E_2(t).
\ee
Therefore, the equation for $n=1$ is given by
\bb
\diffab{E_1(t)}{t}=2\Gamma \Big [
1-2E_1(t)+E_2(t) \Big ].
\ee
In order to be able to write this as the extension of eq.~\eref{E1disc} for $n=1$,
it appears convenient and is, indeed, common, to introduce the constraint $E_0(t)=1$.
We shall do the same, but return to this condition below. However, the boundary conditions, including
$E(0,t)=1$ were considered to be sufficiently complicated so that an explicit solution of \eref{E1cont}
is usually avoided. Ingenious ways have been developed to extract physically interesting information,
such as the particle-density $c(t)$. We shall require the explicit form of $E(x,t)$
below when looking for correlation functions and shall now give it.
In the continuum limit, when $a$ is small, we set $x=n{\rm a}$ and
$E(x,t)=E_n(t)$.
The previous relation \eref{E1disc} can be expanded with respect to $a$ and a rescaled hopping rate $D=\Gamma/a^2$,
which leads to a simple diffusion equation, together with a boundary condition
\bb\label{E1cont}
\diffab{E(x,t)}{t}=2D\diffabc{E(x,t)}{x}{x}, \mbox{\rm ~~and~~} E(0,t)=1.
\ee
If we could use a spatially infinite Fourier transform
$E(x,t)=\int_{-\infty}^{+\infty} \frac{\D k}{2\pi} \exp(\II kx)\wit{E}(k,t)$
to solve the previous equation, we would obtain in the standard fashion
\bb\nn
E(x,t)&=&\int_{-\infty}^{\infty} \!\frac{\D x'}{\sqrt{\pi\,}\,\Lo}\: \exp\Big
[-\frac{1}{\Lo^2}(x-x')^2\Big ]E(x',0), \label{intE1}
\ee
where the integrals over the real axis are unrestricted.
In the above expression, a {\em diffusion length}
\bb
\Lo :=\sqrt{8Dt\,}
\ee
acts as the scaling length of the function $E(x,t)=E(x/\Lo)$.

\subsection{Effect of the boundary condition: continuum limit}

The simplistic approach outlined at the end of the previous subsection must evidently be modified
in order to take the boundary condition $E(0,t)=1$ into account.
This amounts to define in eq.~(\ref{intE1}) the meaning
of the probability $E(x',0)$ for negative $x'$ and is achieved by the following result.

\begin{lemma}
If one extends the validity of eq.~(\ref{E1disc}) to all $n\in\mathbb{Z}$, together with the boundary condition 
$E_0(t)=1$, one has
\bb \label{en}
E_{-n}(t)=2-E_n(t).
\ee
In the continuum limit, this leads to $E(-x,t)=2-E(x,t)$.

\end{lemma}

\begin{proof}
This is proven by induction. First, we consider the case $n=0$.
Using eq.~(\ref{E1disc}) and $E_0(t)=1$, we obtain
\bb
\diffab{E_0(t)}{t}=
2\Gamma
\left (
E_{-1}-2E_0+E_{1} \right ), \nonumber
\ee
which implies $E_{-1}(t)=2E_0(t)-E_{1}(t)=2-E_1(t)$. In the general case, let us
consider the equation of motion for the index $-n-1$ and use the assumption
(\ref{en}) for the indices $-n$ and $-n+1$:
\begin{eqnarray}
E_{-n-1} &=& 2E_{-n} - E_{-n+1} + \frac{1}{2\Gamma} \partial_t E_{-n} \nonumber \\
&=& 2(2-E_n) - (2-E_{n-1}) + \frac{1}{2\Gamma} \partial_t (2-E_{n}) \nonumber \\
&=& 2 -E_{n+1} \nonumber
\end{eqnarray}
where the equations of motion (\ref{E1disc}) were used again. This completes the proof.  \hfill q.e.d.
\end{proof}

In the continuum limit, this relation allows us to rewrite the integral
\eref{intE1} over the positive axis only
\bb\label{E1gen}
\hspace{-1.5truecm}
E(x,t)=\erfc(x/\Lo)+\int_0^{\infty}\!\frac{\D x'}{\sqrt{\pi\,}\,\Lo}\:
E(x',0)
\Big [
e^{-(x-x')^2/\Lo^2}-e^{-(x+x')^2/\Lo^2}
\Big ].
\ee
and where erfc is the complementary error function \cite{Abra}.

Eq.~(\ref{E1gen}) is the general solution for the probability $E(x,t)$ of having
an empty interval, at least of length $x$ and at time $t$,
where the initial state is described by the function $E(x,0)$.
The particle concentration $\conc(t)=\pr(\bullet)/a$ can be obtained in the
continuum limit from the relation \eref{E1}:
\bb\nn
\pr(\bullet)&+&\pr(\circ)=1\;\Rightarrow\pr(\bullet)=a\conc(t)=1-E_1(t),
\ee
where $a\conc(t)=1-E_1(t)\simeq 1-E(0,t)-a\diffab{E(x=0,t)}{x}$, and therefore
\bb\label{concent}
\conc(t)=-\left.\diffab{E(x,t)}{x}\right|_{x=0}.
\ee
The function $E(x,t)$ can by definition be written as a cumulative sum of
the probabilities for having bounded on the left, of size at least equal to $x'$ or 
$P(x',t)=\pr(\bullet\,\fbox{ $x'$ }\,)$:
\bb\label{EP}
E(x,t)=\int_x^{\infty} \!\D x'\, P(x',t).
\ee
This imposes two boundary conditions: first, we have $E(0,t)=\int_0^{\infty}
\!\D x P(x,t)=1$ by normalisation. Then, in the limit $x\to\infty$, one must
have $E(x,t)\to 0$.

We can express $E(x,t)$ as function of $P(x,0)$ by performing an integration by
parts of  \eref{E1gen}:
\bb
E(x,t)=1-\frac{1}{2}\int_0^{\infty} \!\D x'\,P(x',0)\Big [
\erf\Big (\frac{x'+x}{\Lo}\Big )-\erf\Big (\frac{x'-x}{\Lo}\Big )
\Big ].
\ee
By differentiation with respect to $x$, we obtain the expression
for the concentration
\bb\label{concentratGen}
c(t)&=&\frac{2}{\sqrt{\pi}\Lo}\int_0^{\infty} \!\D x'\,
P(x',0)\exp\Big (-\frac{x'^2}{\Lo^2}\Big )
\ee

{}From this, {\em all} initial conditions, characterised by $P(x,0)$, lead to the long-time
behaviour of the concentration:
\begin{lemma}
For sufficiently long times and any initial distribution $P(x,0)$,  the concentration decreases as
\bb \label{gl:lemma2.2}
c(t) \simeq \frac{2}{\sqrt{\pi\,}}\frac{1}{\Lo} + {\rm o}\left( \Lo^{-1}\right)
\ee
\end{lemma}
\begin{proof}
Since $P(x,0)$ is a normalised probability distribution, $\int_0^{\infty} \! \D x\, P(x,0)=1$, we must have
$P(x,0)={\rm o}(1/x)$ for $x\to\infty$. We rewrite eq.~(\ref{concentratGen}) as follows
\begin{eqnarray}
c(t) &=&\frac{2}{\sqrt{\pi\,}}\frac{1}{\Lo} -
\frac{2}{\sqrt{\pi\,}} \underbrace{\int_0^{\infty} \! \D x\, P(x\, \Lo,0) \left( 1 - e^{-x^2} \right)}_{{\rm o}(1/\Lo)}
\end{eqnarray}
where the last estimate follows from the large-$x$ behaviour of $P(x,0)$. \hfill q.e.d.
\end{proof}
In a more explicit way, this may be obtained from eq.~(\ref{concentratGen}) by a formal expansion of the
exponential. If the second moment of $P$ is well-defined, this leads to the long-time behaviour
\bb
\conc(t)\simeq \frac{2}{\sqrt{\pi}\Lo}\Big (1-\frac{\langle
x^2\rangle }{\Lo^2}\Big )+\mbox{\rm o}(\Lo^{-2}).
\ee
On the other hand, the case of a diverging second moment is illustrated by the example 
\bb
P(x,0)= \frac{a_0}{1+x^{1+\alpha}},
\ee
where $0<\alpha<1$ and
$a_0=(1+\alpha)\sin[\pi/(1+\alpha)]/\pi$ is the normalisation factor.
In this case, a calculation analogous to the proof of lemma 2.2 gives
\bb\nn
\hspace{-1.5truecm}c(t)&=&\frac{2}{\sqrt{\pi\,}\,\Lo}\int_0^{\infty}
\!\D x\,a_0\frac{\exp\left(-\frac{x^2}{\Lo^2}\right)}{1+x^{1+\alpha}}
\\
\hspace{-1.5truecm}&=&
\frac{2}{\sqrt{\pi\,}\,\Lo}
\left[
1-\frac{a_0}{\Lo^{\alpha}}\int_0^{\infty}\!\D x\,\frac{1-e^{-x^2}}{
x^{ 1+\alpha } }
+\cdots \right],
\ee

which gives the leading correction in the long-time limit as function of the
exponent $\alpha$.
We also notice that if particles occupy each site with probability $p$, the
concentration is defined by 
$c_0=p/a$. When the
system is filled with a concentration $c_0$ of particles, the function
$E_n(0)$ is proportional to
\bb
E_n(0)\sim(1-p)^{n}=(1-ac_0)^{x/a}
\stackrel{a \rightarrow 0}{\longrightarrow}\;E(x,0)=\exp(-c_0x)
\ee
However, if the system is entirely filled with particles, $p=1$ and
$E_n(0)=0$ for $n\neq0$, then, from \eref{E1gen}, $E(x,t)$ is simply given by $\erfc(x/\Lo)$,
and $c(t)= 2/\sqrt{\pi}\Lo$. In the general case of a given concentration $
c_0$ where $E_0(x)=e^{-c_0x}$, we simply
have $P(x,0)=c_0e^{-c_0x}$ and from \eref{concentratGen}
\bb \label{eq:ccont}
c(t)=c_0\exp\left(\frac{1}{4} c_0^2\Lo^2\right)\,\erfc\left(\frac{c_0\Lo}{2}\right).
\ee

\begin{figure}[th]
\centering\resizebox{0.8\columnwidth}{!}
{\includegraphics*{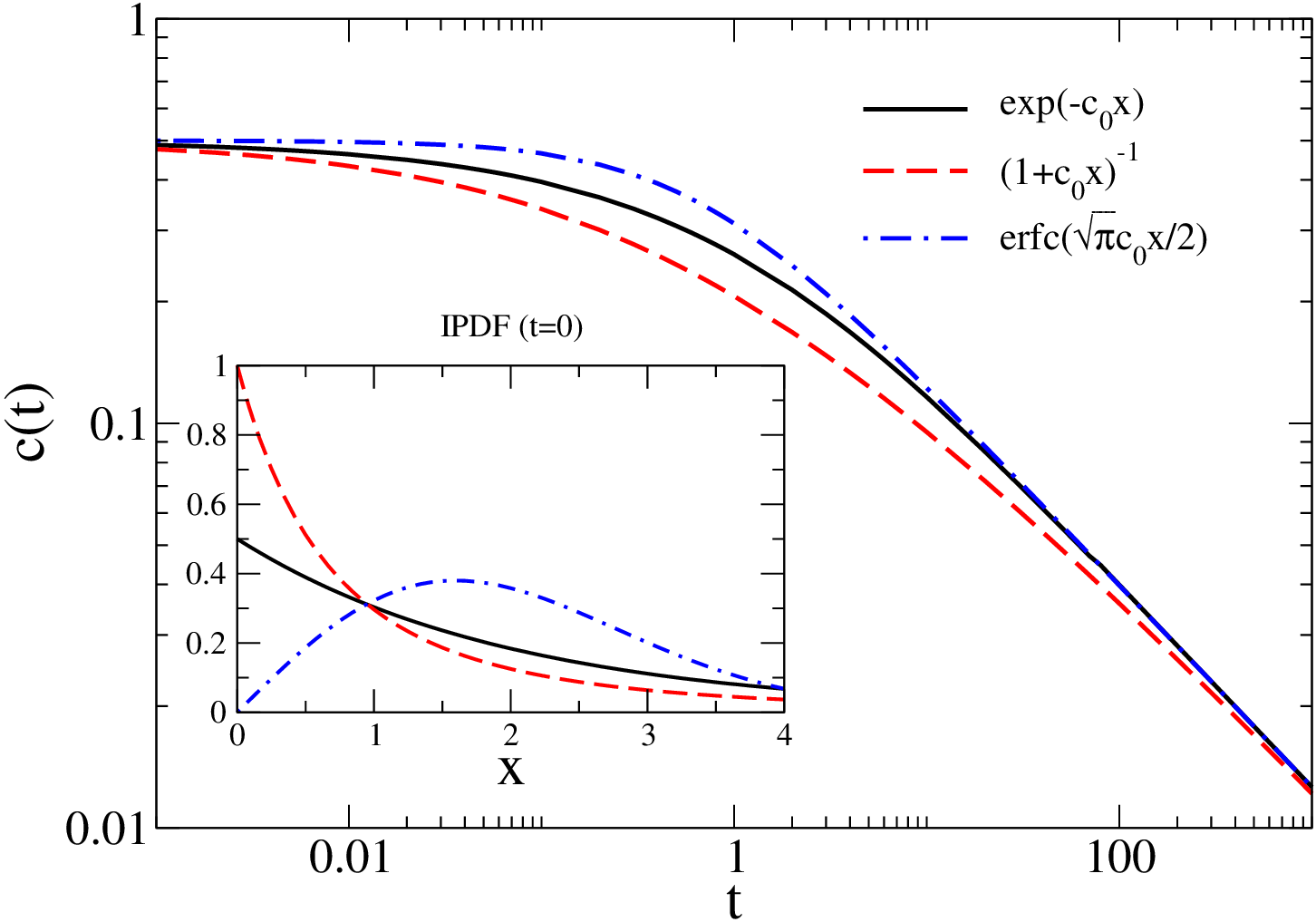}}
\caption{\label{fig1}
Time-evolution of the concentration $c(t)$, for several initial conditions expressed in terms of $E(x,0)$ and parameter 
$c_0=1/2$. For any initial distribution, the particle concentration shows the same asymptotic behaviour.
The passage between the initial and the asymptotic regime can be qualitatively explained
in terms of the interparticle distribution function (IPDF) defined in eq.~(\ref{eq:ipdf}) and shown for $t=0$ in the 
inset. See text for details.}
\end{figure}

In figure~\ref{fig1} we illustrate the effect of several initial empty-interval distributions $E(x,0)$.
Clearly, as expected from the above discussion, all initial distributions lead to the same long-time asymptotics 
$c(t) \sim t^{-1/2}$ but the way this asymptotic regime is reached depends on the initial state. 
This can be better understood when plotting the interparticle distribution function (IPDF)
\bb \label{eq:ipdf}
p(x,t) := \frac{1}{c(t)}\frac{\partial^2 E(x,t)}{\partial x^2},
\ee
which gives the probability density that the next neighbour of a particle is at distance
$x$ at time $t$ \cite{benA90,benA00}. This function is shown for $t=0$ in the inset of figure~\ref{fig1}.
One observes that in those cases when $p(x,0)$ decays monotonously with $x$, the transition to the
asymptotic regime is more gradual. On the other hand, in the third case there is an initial non-vanishing distance the
particles must overcome before they can react. This leads to a very sharp transition between the initial
and the asymptotic regimes.

\subsection{The discrete case}

We now give the solution of the discrete case, without performing a continuum limit.

\begin{figure}
\centering\resizebox{0.8\columnwidth}{!}
{\includegraphics*{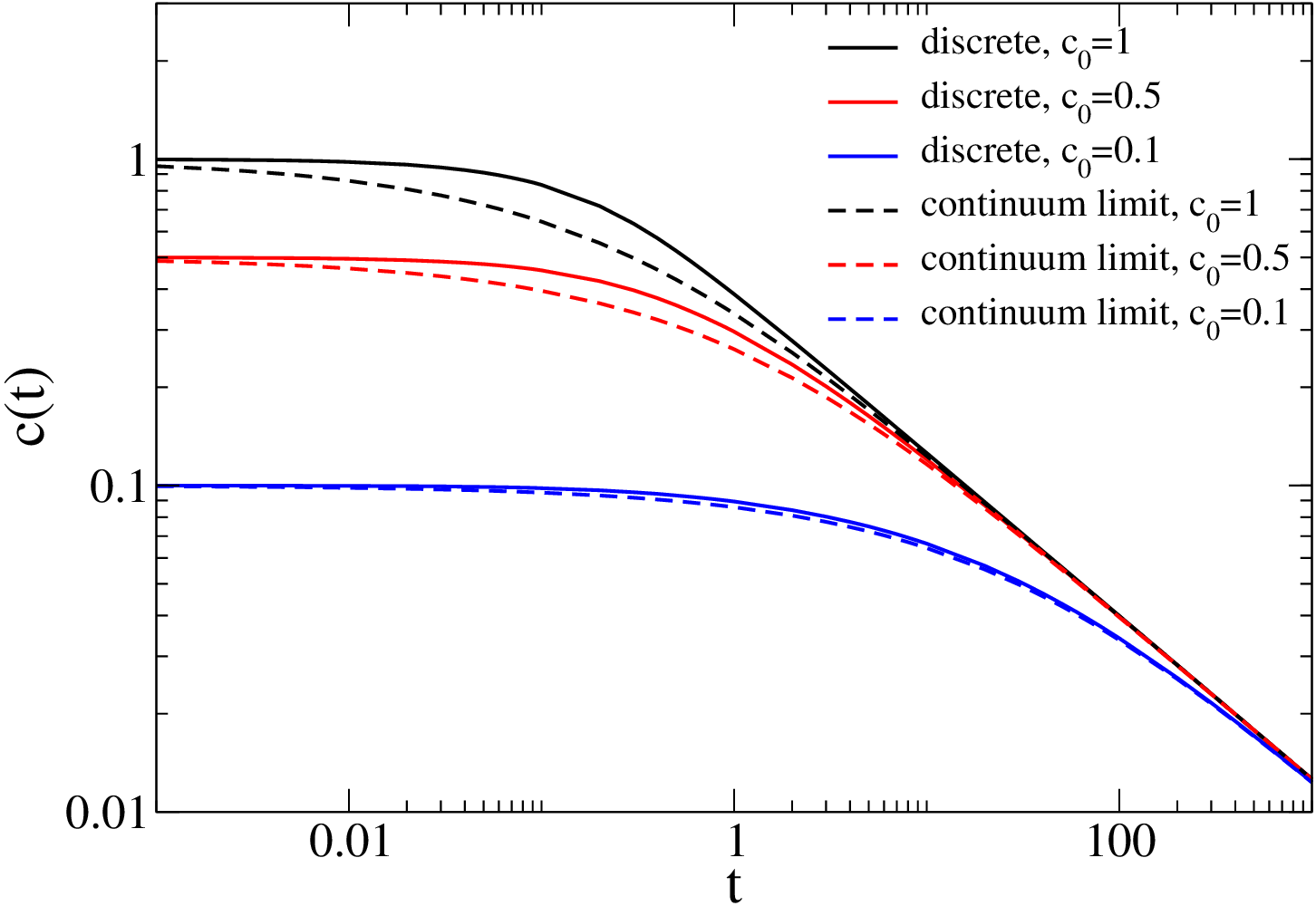}}
\caption{\label{fig2}
Time evolution of the concentration in the discrete case (full curves) and the continuous limit (dashed lines). 
The initial concentration $c_0$ is $[1,0.5,0.1]$ from top to bottom.}
\end{figure}

First we recall the solution of the differential equation (\ref{E1disc}).  The generating function 
$F(z,t) = \sum_{n=-\infty}^{+\infty} z^n E_n(t)$ satisfies as usual the
differential equation $\partial_t F(z,t) = 2D(z+1/z-2) F(z,t)$, with the solution
\bb
\hspace{-1.5cm}F(z,t) = F(z,0)e^{2D(z+1/z-2)t}
= e^{-4Dt}\sum_ {n=-\infty}^{+\infty}z^n\sum_ {m=-\infty}^{+\infty} E_{n-m}(0)I_m(4Dt)
\ee
Identifying $E_n(t)$ in the previous expression, we write
\bb
E_n(t)=e^{-4Dt}\sum_ {m=-\infty}^{+\infty} E_m(0)I_{n-m}(4Dt).
\ee
We now must take the boundary condition $E_0(t)=1$ into account.
As in the continuum case, we replace the summation over negative values of the index $m$ by using the discrete relation 
$E_{-n}(t)=2-E_n(t)$ and find
\BEQ
\hspace{-2.45cm}
E_n(t) = e^{-4Dt}\left[\sum_
{m=1}^{+\infty}E_m(0)\Bigl(I_{n-m}(4Dt)-I_{n+m}(4Dt)\Bigr) + \!\sum_
{m=1}^{+\infty} 2I_{n+m}(4Dt) + I_1(4Dt)\right].
\EEQ
In the discrete case, the particle concentration is given by $c(t) = 1-E_1(t)$. 
Using summation and recurrence relations over modified Bessel functions, we obtain
\bb \label{eq:cdisc}
c(t) = e^{-4Dt}\left(I_0(4Dt)+I_1(4Dt) - \sum_ {m=1}^{+\infty}
\frac{m}{2Dt}E_m(0)I_m(4Dt)\right),
\ee
which generalises earlier results of Spouge \cite{Spouge88}. In figure \ref{fig2}, we compare the particle concentration 
according to the discrete case eq.~(\ref{eq:cdisc}) with the previously obtained solution eq.~(\ref{eq:ccont}) in the 
continuum limit, for three values of the initial concentration $c_0$. We used, respectively, the initial distributions 
$E_n(0)=(1-c_0/a)^n$ and $E(x,0)=e^{-c_0 x}$. As expected, the same asymptotics is found in all cases, independently of 
the initial concentration. We observe that the passage between the initial and the asymptotic regimes is more gradual in 
the continuum limit. As above, we interpret this as coming from the fact that in the discrete case the particles must 
first overcome a finite distance before then can react.

\section{Two-interval probability}

In this section, we generalize the previous result and evaluate
the probability $E_{n_1,n_2}(d,t)$ to have {\em two} empty intervals, at least of sizes $n_1$
and $n_2$ and separated by the distance $d$: we denote it by $E_{n_1,n_2}(d,t)=\pr\bigl(\fbox{ $n_1$ }
\; d \;\fbox{ $n_2$ }\,\bigr)$. This function is expected to have the
following symmetries
\bb\nn
& &E_{n_1,n_2}(d,t)=E_{n_2,n_1}(d,t)
\\ \nn
& &E_{n_1,0}(d,t)=E_{n_1}(t) \quad \mbox{and} \quad E_{0,n_2}(d,t)=E_{n_2}(t)
\\ \label{E2sym}
& &E_{n_1,n_2}(0,t)=E_{n_1+n_2}(t).
\ee

\subsection{Equations of motion}

As before and using the notations of previous section, we consider the different possibilities for
the variation of $E_{n_1,n_2}(d,t)$ between the time $t$ and $t+\D t$:
\bb
\hspace{-2cm}
\diffab{E_{n_1,n_2}(d,t)}{t}\D t&=&
-\left[
\pr(\bullet\rar\fbox{ $n_1$ }\;d\;\fbox{ $n_2$ })
+
\pr(\fbox{ $n_1$ }\;d\;\fbox{ $n_2$ }\lar\bullet)\right.
\\ \nn
& & \quad +\left.
\pr(\fbox{ $n_1$ }\lar\bullet d-1\;\fbox{ $n_2$ })
+
\pr(\fbox{ $n_1$ }\;d-1\bullet\rar\fbox{ $n_2$ })
\right]
\\ \nn
& &+\left[
\pr(\lar\bullet\fbox{ $n_1-1$ }\;d\;\fbox{ $n_2$ })
+
\pr(\fbox{ $n_1$ }\;d\;\fbox{ $n_2-1$ }\bullet\rar) \right.
\\ \nn
& & \quad +\left.
\pr(\fbox{ $n_1-1$ }\bullet\rar d\;\fbox{ $n_2$ })
+
\pr(\fbox{ $n_1$ }\;d\lar\bullet\fbox{ $n_2-1$ })
\right].
\ee
The probability rates are given by considering the sum rules for
static probabilities. First, we consider the negative contributions for which we obtain the
relations
\bb\nn
\hspace{-2cm}& &\pr(\bullet\rar\fbox{ $n_1$ }\;d\;\fbox{ $n_2$ })=\pr(\bullet\fbox{ $n_1$
}\;d\;\fbox{ $n_2$ })\,\Gamma \D t,
\\ \nn
& &\pr(\bullet\fbox{ $n_1$}\;d\;\fbox{ $n_2$ })+\pr(\circ\fbox{
$n_1$}\;d\;\fbox{ $n_2$ })=\pr(\fbox{ $n_1$ }\;d\;\fbox{ $n_2$ })
\\ \nn
& &\Rightarrow
\pr(\bullet\fbox{ $n_1$ }\;d\;\fbox{ $n_2$ })=E_{n_1,n_2}(d,t)-E_{n_1+1,n_2}(d,t)
\ee
and
\bb \nn
\hspace{-2cm}& &\pr(\fbox{ $n_1$ }\lar\bullet d-1\;\fbox{ $n_2$ })
=\pr(\fbox{ $n_1$ }\bullet d-1\;\fbox{ $n_2$ })\,\Gamma \D t,
\\ \nn
& &\pr(\fbox{ $n_1$ }\bullet d-1\;\fbox{ $n_2$ })+
\pr(\fbox{ $n_1$ }\circ d-1\;\fbox{ $n_2$ })=
\pr(\fbox{ $n_1$ } d\;\fbox{ $n_2$ })
\\ \nn
& &\Rightarrow
\pr(\fbox{ $n_1$ }\bullet d-1\;\fbox{ $n_2$ })
=
E_{n_1,n_2}(d,t)-E_{n_1+1,n_2}(d-1,t).
\ee
For the positive contibutions, we have
\bb\nn
\hspace{-2cm}\pr(\lar\bullet\fbox{ $n_1-1$ }\;d\;\fbox{ $n_2$ })=
\pr(\bullet\fbox{ $n_1-1$ }\;d\;\fbox{ $n_2$ })\,\Gamma \D t,
\\ \nn
\hspace{-2cm}\pr(\bullet\fbox{ $n_1-1$ }\;d\;\fbox{ $n_2$ })+
\pr(\circ\fbox{ $n_1-1$ }\;d\;\fbox{ $n_2$ })=
\pr(\fbox{ $n_1-1$ }\;d\;\fbox{ $n_2$ })
\\ \nn
\Rightarrow
\hspace{-2cm}\pr(\bullet\fbox{ $n_1-1$ }\;d\;\fbox{ $n_2$
})=E_{n_1-1,n_2}(d,t)-E_{n_1,n_2}(d,t),
\ee
and
\bb\nn
\hspace{-1cm}& &\hspace{-1.5cm}\pr(\fbox{ $n_1-1$ }\bullet\rar d\;\fbox{ $n_2$ })=
\pr(\fbox{ $n_1-1$ }\bullet d\;\fbox{ $n_2$ })\,\Gamma \D t,
\\ \nn
& &\hspace{-1.5cm}\pr(\fbox{ $n_1-1$ }\bullet d\;\fbox{ $n_2$ })+
\pr(\fbox{ $n_1-1$ }\circ d\;\fbox{ $n_2$ })=
\pr(\fbox{ $n_1-1$ }\;d+1\;\fbox{ $n_2$ })
\\ \nn
& &\hspace{-1.5cm}\Rightarrow
\pr(\fbox{ $n_1-1$ }\bullet d\;\fbox{ $n_2$ })=
E_{n_1-1,n_2}(d+1,t)-E_{n_1,n_2}(d,t),
\ee
(similarly for the other terms which are symmetric). After
gathering all the contributions, we finally find (the time variable is from now on suppressed)
\bb\label{E2disc}
\hspace{-1.5cm}
\lefteqn{\diffab{E_{n_1,n_2}(d)}{t} = \Gamma \left[ -8E_{n_1,n_2}(d)\right.}
\\ \nn
& &\hspace{-1.0truecm}+
E_{n_1+1,n_2}(d)+E_{n_1,n_2+1}(d) + E_{n_1-1,n_2}(d)+E_{n_1,n_2-1}(d)
\\ \nn
& &\hspace{-1.0truecm}+\left.
E_{n_1+1,n_2}(d-1)+ E_{n_1,n_2+1} (d-1)+E_{n_1-1,n_2}(d+1)+E_{n_1,n_2-1}(d+1)
\right].
\ee
We have checked that the same closed system of equations of motion is also obtained when the master equation is rewritten in terms of a quantum Hamiltonian \cite{pesch}.

The continuum limit of this diffusion equation is obtained by expanding the
terms up to the second order in the lattice step $a$ (the $a^2$ will be absorbed
in $\Gamma$). Setting $x=n_1a$, $y=n_2a$ and $z=d\,a$ we obtain
the following linear differential equation:
\bb\label{E2cont}
\diffab{E(x,y,z)}{t}=2D
\Big [
\partial_x^2+\partial_y^2+\partial_z^2-
\Big (
\partial_{x}\partial_z+\partial_y\partial_z
\Big )
\Big ]E(x,y,z).
\ee

\subsection{General solution}

The general solution for \eref{E2cont} is obtained by diagonalising
the quadratic form associated with the differential operator
$\wht{P}:=\partial_x^2+\partial_y^2+\partial_z^2-\partial_{x}
\partial_z-\partial_y\partial_z$. We find that the following change of variables
$(x,y,z)\rightarrow(X,Y,Z)$ diagonalises $\hat P$:
\BEQ \label{XYZ}
X = \alpha(x+y+\sqrt{2}z) \;\; ,\;\;
Y = \beta(x+y-\sqrt{2}z) \;\; , \;\;
Z = \gamma (x-y),
\EEQ
with the positive constants
\bb
\alpha&=&[2(2-\sqrt{2})]^{-1/2} \;\;,\;\; \beta=[2(2+\sqrt{2})]^{-1/2}
\;\;,\;\; \gamma=1/\sqrt{2}
, \nn \label{constants}
\ee
such that $\alpha^2+\beta^2=1$ and $\alpha^2-\beta^2=1/\sqrt{2}$.

In the new variables, the operator
$\wht{P}=\partial_X^2+\partial_Y^2+\partial_Z^2$ is diagonal.
As in the previous section (see \eref{intE1}) for the one-interval problem, and
in the continuum limit, if the boundary conditions are ignored for a moment,
the function $E(x,y,z)$ can be found via a Fourier transformation, and explicitly
expressed as a kernel integral depending on the initial conditions
$E_0(x,y,z):= E(x,y,z,0)$ :
\bb\label{E2gen}
\hspace{-1.5truecm}E(x,y,z)=\frac{\sqrt{2}}{(\sqrt{\pi}\,\Lo)^3}\,\int_{-\infty
} ^ { \infty } \!\D x'\D y'\D z'\:
\wesl{x-x',y-y',z-z'}
E_0(x',y',z')
\ee
where the Gaussian kernel  $\wesl{u,v,w}$ is given by
\bb\nn
\hspace{-1.8truecm}\wesl{u,v,w}&=&\exp\frac{1}{\Lo^2}\Big [
-\alpha^2(u+v+\sqrt{2}w)^2-\beta^2(u+v-\sqrt{2}w)^2-\gamma^2(u-v)^2
\Big ]
\\ \label{kernel}
&=&
\exp
\frac{1}{\Lo^2}\Big [
-(u+v+w)^2-w^2-\frac{1}{2}(u-v)^2
\Big ],
\ee
the Jacobian of the transformation \eref{XYZ} being equal to
$4\sqrt{2}\alpha\beta\gamma=\sqrt{2}$.

\subsection{Compatibility conditions}

As expressed before by equation \eref{en}, it is important to make a
correspondence between intervals of formally negative
and positive lengths. This will be needed in the formal solution (\ref{E2gen}) which requires real variables,
whereas the probability $E(x,y,z)$ has an obvious  physical meaning depends for positive distances only.
We have to consider 3 cases, depending on whether $x$, $y$ or $z$ are negative or positive.
By symmetry considerations \eref{E2sym}, it is only necessary to consider the case where
$x$ or $z$ are
individually negative (the case $y<0$ being deduced by means of the first equation \eref{E2sym}), 
the other variables being positive. The explicit evaluation of eq.~(\ref{E2gen}) in the following sections requires  
several identities, stated as lemmata for clarity and proven in appendix~A and~B, respectively. The first one
treats the case of formally negative interval lengths. 

\begin{lemma}
The probability of two-empty-intervals of negative lengths is related to the probability of
positive lengths as follows.
\bb\nn
E_{-n_1,n_2}(d)=2E_{n_2}-E_{n_1,n_2}(d-n_1),
\\ \label{E2n12}
E_{n_1,-n_2}(d)=2E_{n_1}-E_{n_1,n_2}(d-n_2),
\\ \label{E2n123}
E_{-n_1,-n_2}(d)=4-2E_{n_1}-2E_{n_2}+E_{n_1,n_2}(d-n_1-n_2).
\ee
\end{lemma}
A further relation connects the negative separations $-d$ between two intervals to the positive ones.
\begin{lemma}
We have
\bb\label{E2d}
E_{n_1,n_2}(-d)=2E_{n_1+n_2-d}-E_{n_1-d,n_2-d}(d).
\ee
\end{lemma}

Later on, we shall require these results in the continuum limit, where the expressions 
\eref{en}, \eref{E2n12}, \eref{E2n123} and \eref{E2d} take the following form
\bb\nn
E(-x)=2-E(x),
\\ \nn
E(-x,y,z)=2E(y)-E(x,y,z-x),
\\ \nn
E(x,-y,z)=2E(x)-E(x,y,z-y),
\\ \nn
E(-x,-y,z)=4-2E(x)-2E(y)+E(x,y,z-x-y),
\\
E(x,y,-z)=2E(x+y-z)-E(x-z,y-z,z).
\ee
This allows us to rewrite
\eref{E2gen} in the restricted domain where $(x',y',z')$ are all positive,
and where $E_0(x',y',z')$ is physically well-defined.

\section{General solution for $E(x,y,z,t)$}

{}From the general equation \eref{E2gen}, we separate the 8 different domains of
integration around the origin for example $(x'>0,y'>0,z'>0)$,
$(x'<0,y'>0,z'>0)$ etc.., and use relations \eref{en},
\eref{E2n12}, \eref{E2n123} and
\eref{E2d} to map all domains into the single domain $(x'>0,y'>0,z'>0)$. This calculation is done in the appendix C and, here, we just summarize the results. 
The general solution can be decomposed as follows:
\bb \label{gl:E2deco}
E(x,y,z,t)=E^{(0)}(x,y,z,t)+E^{(1)}(x,y,z,t)+E^{(2)}(x,y,z,t) ,
\ee
where $E^{(0)}(x,y,z,t)$ is obtained from the terms independent of the initial conditions,
$E^{(1)}(x,y,z,t)$ from the initial one-interval probability $E_0(x')$ and
$E^{(2)}(x,y,z,t)$ from the initial two-interval probability $E_0(x,y,z)$,
respectively.
Note that $E^{(1)}(x,y,z,t)$ and $E^{(2)}(x,y,z,t)$ depend on $E_0(x')$, $E_0(x',y',z')$ with arguments positive, hence this gives us the physical answer to the diffusion process in the
coagulation problem starting from arbitrary initial conditions and constraints
on the differential equation.
We now analyse these three terms one by one.

\subsection{Special case of a system initially entirely filled with particles}

We notice that \eref{partx}, \eref{party}, and \eref{partxy}
contain initial conditions for the single-interval distribution $E_0(x')$, some
constants independent of the initial conditions, and initial conditions for the
two-interval distribution $E_0(x',y',z')$. To simplify notations, we shall
re-scale all lengths by $\Lo$ such that $E_{\ell_0}(x,y,z)=E(x \Lo,y \Lo,z
\Lo,t)$.
{}In \eref{partxy}, we can isolate from $E_0(-x',-y',z')$ two terms
independent of the initial conditions, 
\bb\nn
4-4\theta(y'-z')\theta(x'-z')\theta(x'+y'-z').
\ee
It is obvious that $\theta(y'-z')\theta(x'-z')\theta(x'+y'-z')=\theta(y'-z')\theta(x'-z')$.
These two terms, plus the first term in \eref{E2va}, give a contribution to the
general function equal to
\bb \nn
\hspace{-1.5truecm}E^{(0)}_\Lo(x,y,z)=
\erfc(z)\erfc(x+y+z)
\\ \nn
\hspace{-1.5truecm}+\sqrt{\frac{2}{\pi^3}\,}
\int_{\mathbb{R}_+^3} \!\D x'\D y'\D z'\:
\wet{-x',-y',z'}\{4-4\theta(y'-z')\theta(x'-z')\}.
\ee
Performing the translations $x'-z'\rightarrow x'$ and $y'-z'\rightarrow y'$ in the last
contribution, we obtain
\bb \nn
E^{(0)}_\Lo(x,y,z)=
\erfc(z)\erfc(x+y+z)
\\ \nn
+\sqrt{\frac{2}{\pi^3}\,}
\int_{\mathbb{R}_+^3} \!\D x'\D y'\D z'\:
\Bigl\{
\wet{-x',-y',z'}-\wet{-x'-z',-y'-z',z'}
\Bigr\}.
\ee
Using the relation \eref{relation-z}, and the identity
\bb\nn
\we{x-x'-z',y-y'-z',z+z'}=\we{x-x',y-y',z-z'}e^{-4zz'}
\\ \label{relation-z2}
\Rightarrow \wet{x'+z',y'+z',-z'}=-\wet{x',y',z'}
\ee
we can rewrite $E$ as
\bb\nn
\hspace{-1.5truecm}E^{(0)}_\Lo(x,y,z)&=&
\erfc(z)\erfc(x+y+z)
\\ \nn
& &+\sqrt{ \frac{32} {\pi^3}\,}
\int_{\mathbb{R}_+^3} \!\D x'\D y'\D z'\:
\Bigl\{
\wet{-x',-y',z'}+\wet{-x',-y',-z'}
\Bigr\}
\\ \nn
&=&
\erfc(z)\erfc(x+y+z)
\\ \nn
& &+\sqrt{ \frac{32} {\pi^3}\,}
\int_{\mathbb{R}_+^2} \!\D x'\D y' \int_{\mathbb{R}} \! \D z'\:
\wet{-x',-y',z'}
\ee
The integral over $z'$ now gives a gaussian
exponential. Then the two remaining integrals can also be carried out
explicitly (see appendix~G). Introducing again the diffusion
length $\Lo$, we find
\bb\nn
E^{(0)}(x,y,z)&=&
\erfc\left(\frac{x}{\Lo}\right)\erfc\left(\frac{y}{\Lo}\right)
+\erfc\left(\frac{z}{\Lo}\right)\erfc\left(\frac{x+y+z}{\Lo}\right)
\\ \label{E2lim}
& &-\erfc\left(\frac{x+z}{\Lo}\right)\erfc\left(\frac{y+z}{\Lo}\right).
\ee
This is the exact two-interval probability in the case of an initially fully filled lattice (where
both $E_0(x)$ and $E_0(x,y,z)$ vanish). In particular, the solution (\ref{E2lim}) satisfies the symmetry conditions
\eref{E2sym}. In the limit of $z$ large, one has the factorisation
$E(x,y,z,t)\simeq E(x,t)E(y,t)$.

The remaining terms of the full solution depend on the initial conditions. 
They are of two kinds, and involve either
the single-interval or else the two-interval initial probabilities. We turn to them now.

\subsection{Contributions to $E(x,y,z,t)$ from terms with a single-interval initial distribution}

The contributions to $E(x,y,z,t)$ of single-interval distributions come from the
previous relations \eref{partx}, \eref{party}
and \eref{partxy}, where we can isolate the following individual terms
\begin{itemize}
\item the second term of equation \eref{E2va}
\item the first 3 terms in \eref{partx} and \eref{party}
\item terms 2, 3, 4 in \eref{partxy}.
\end{itemize}
On the whole, there are 10 terms contributing to the initial conditions given
for a given choice of $E_0(x')$. Gathering these terms and performing successive
translations in $x'$, $y'$ or $z'$ when necessary, we obtain
\bb\nn
\hspace{-1cm}
E^{(1)}_\Lo(x,y,z)=
\erfc(z)\int_0^{\infty}
\frac{\D x'}{\sqrt{\pi\,}}\: E_{0,\Lo}(x')
\Big (
e^{-(x+y+z-x')^2}-e^{-(x+y+z+x')^2}
\Big )
\\ \nn
\hspace{-1.5cm}+\sqrt{\frac{8}{\pi^3}\,} \int_{\mathbb{R}_+^3} \!\D x'\D y'\D z'\;
E_0(x')
\Big \{
\underbrace{\wet{x',-y',z'}-\wet{x'-z',-y'-z',z'}}_{I_1}
\\ \nn
\hspace{-1.5cm}+\underbrace{\wet{-y',x',z'}-\wet{-y'-z',x'-z',z'}}_{I_2}
\\ \nn
\hspace{-1.5cm}+\underbrace{\wet{-x'-z',-y'-z',z'}-\wet{-x',-y',z'}}_{I_3}
\\ \nn
\hspace{-1.5cm}+\underbrace{\wet{-y'-z',-x'-z',z'}-\wet{-y',-x',z'}}_{I_4}
\\ \nn
\hspace{-1.5cm}+\underbrace{\theta(-x'-y'+z')[\wet{y',-z',x'}+\wet{-z',y',x'}]}_{I_5}
\\ \label{E21a}
\hspace{-1.5cm}+
\underbrace{\theta(-x'+y'+z')\wet{-z',-y',x'}}_{I_6}
\Big \}
\ee
By performing partial translations and integrations and simplifying all terms
from $I_1$ to $I_6$ (see appendix~D for details), we obtain
\bb\nn
\hspace{-2cm}E^{(1)}_\Lo(x,y,z)=
\int_0^{\infty}
\frac{\D x'}{\sqrt{\pi\,}}\: E_{0,\Lo}(x')
\Big \{
\erfc(z)\Big [
e^{-(x'-x-y-z)^2}-e^{-(x'+x+y+z)^2}
\Big ]
\\ \nn
\hspace{-1.5cm}+
\erfc(x)
\Big [
e^{-(x'-y)^2}-e^{-(x'+y)^2}
\Big]
-\erfc(x+z)
\Big [
e^{-(x'-y-z)^2}-e^{-(x'+y+z)^2}
\Big ]
\\ \nn
\hspace{-1.5cm}+
\erfc(y)
\Big [
e^{-(x'-x)^2}-e^{-(x'+x)^2}
\Big]
-\erfc(y+z)
\Big [
e^{-(x'-x-z)^2}-e^{-(x'+x+z)^2}
\Big ]
\\ \nn
\hspace{-1.5cm}+
\erfc(x+y+z)
\Big [
e^{-(x'-z)^2}-e^{-(x'+z)^2}
\Big ]
\Big \}
\\ \label{E21gen}
\hspace{-1.5cm}=: \int_0^{\infty}
\frac{\D x'}{\sqrt{\pi\,}}\: E_{0,\Lo}(x')K_{1,\Lo}(x';x,y,z),
\ee
where the kernel $K_{1,\Lo}$ is positive. When $z=0$, we recover the result
\eref{E1gen} for a single-interval distribution of size $x+y$. For some
functions $E_0(x')$, the previous integrals can be performed exactly since
$K_1$ is gaussian in the variable $x'$. For example, if we take as initial function
$E_0(x)=e^{-c_0x}$, where $c_0$ is an initial concentration of particles, we find
\BEA 
\nn \hspace{-2.3cm}
\lefteqn{ E^{(1)}(x,y,z)=
\erfc\left(\frac{z}{\Lo}\right)F_{c_0}(x+y+z)
+\erfc\left(\frac{x}{\Lo}\right)F_{c_0}(y)-\erfc\left(\frac{x+z}{\Lo}\right)F_{c_0}(y+z) }
\\ \label{E21c0}
\hspace{-2.0cm}+\erfc\left(\frac{y}{\Lo}\right)F_{c_0}(x)-\erfc\left(\frac{y+z}{\Lo}\right)F_{c_0}(x+z)
+\erfc\left(\frac{x+y+z}{\Lo}\right)F_{c_0}(z),
\EEA
with the following abbreviation
\bb
\hspace{-1.4truecm}F_{c_0}(x) :=
\int_0^{\infty}
\frac{\D x'}{\Lo\sqrt{\pi\,}}\;
e^{-c_0x'}
\Big [
e^{-{(x'-x)^2}/{\Lo^2}}-e^{-{(x'+x)^2}/{\Lo^2}}
\Big]
\\ \nn
\hspace{-1.4truecm}=\frac{1}{2}e^{{\Lo^2 c_0^2}/{4}}
\left\{
e^{-c_0x}\erf\left(\frac{x}{\Lo}-\frac{\Lo c_0}{2}\right)+e^{c_0x}\erf\left(\frac{x}{\Lo}+\frac{\Lo c_0}{2}\right)-2\sinh(c_0x)
\right\}.
\ee
The limiting values of this function read
\bb
\hspace{-1.4truecm}F_{c_0}(x)\stackrel{x\rightarrow \infty}{\simeq}
e^{c_0^2\Lo^2/4-c_0x},\;\;
F_{c_0}(x)\stackrel{x\rightarrow 0}{\simeq}
\Big (
\frac{2}{\Lo\sqrt{\pi}}
-c_0e^{c_0^2\Lo^2/4}\erfc(\Lo c_0/2)
\Big )x,~~~
\ee
and $F_{0}(x)=\erf(x/\Lo)$.

In the long time limit, the above expression goes to zero like
$1/\Lo^3$
\bb
F_{c_0}(x)\simeq \frac{4x}{\sqrt{\pi}c_0^2\Lo^3}
\Big ( 1-\frac{6}{c_0^2\Lo^2}-\frac{x^2}{\Lo^2}+\dots
\Big )
\ee
and the dominant part of contribution $E^{(1)}(x,y,z,t)$ in the same
limit behaves like
\bb\label{E1series}
E^{(1)}(x,y,z)\simeq \frac{4(x+y)}{\sqrt{\pi}c_0^2\Lo^3}.
\ee
It is interesting to compare this expansion with the long-time limit expansion
of $E^{(0)}(x,y,z)$, which is independent of $c_0$

\bb
E^{(0)}(x,y,z)\simeq 1-\frac{2(x+y)}{\sqrt{\pi}\Lo^2}+\frac{2\Big [
(x+y)^3+6xyz \Big ]}{3\sqrt{\pi}\Lo^3}.
\ee
The first term \eref{E1series} tends to increase the two-interval probability
by a factor independent of the distance, since there are less
particles in the system for a finite concentration of particles.

\subsection{Contributions to E(x,y,z) of  two-interval initial distributions}

As noticed previously, we can isolate from equations \eref{partx}, \eref{party},
and
\eref{partxy} terms involving $E(x',y',z',t=0)=E_0(x',y',z')$. In particular

\begin{itemize}
\item 1 term $E_0(x',y',z')$ when all variables are positive, in combination
with $\wet{x',y',z'}$
\item 2$\times$3 terms in \eref{partx} and \eref{party}
\item 5 terms in \eref{partxy}.
\end{itemize}
When combining and simplifying these terms (see appendix~E for details), we
obtain a reduced integral form for $E^{(2)}_\Lo(x,y,z)$, as function of a kernel
$K_{2,\Lo}(x',y',z';x,y,z)$
\bb \nn
\hspace{-2.3truecm}E^{(2)}_\Lo(x,y,z)=\sqrt{\frac{2}{\pi^3}\,}\int_{\mathbb{R}_+^3}
\!\!\D x'\D y'\D z' \: E_{0,\Lo}(x',y',z')\we{x-x',y-y',z-z'}
K_{2,\Lo}(x',y',z';x,y,z)
\ee
with
\bb\nn
K_{2,\Lo}(x',y',z';x,y,z)=
[1-e^{-4(x'+y'+z')(x+y+z)}](1-e^{-4z'z})
\\ \nn
+e^{-4x'x-4y'y}[1-e^{-4z'(x+y+z)}][1-e^{-4(x'+y'+z')z}]
\\ \nn
-e^{-4x'x}[1-e^{-4(y'+z')(x+y+z)}][1-e^{-4(x'+z')z}]
\\ \nn
-e^{-4y'y}[1-e^{-4(x'+z')(x+y+z)}][1-e^{-4(y'+z')z}]
\\ \nn
+e^{-4x'x-4z'(x+z)}[1-e^{-4y'(x+y+z)}](1-e^{-4x'z})
\\ \label{K2}
+e^{-4y'y-4z'(y+z)}[1-e^{-4x'(x+y+z)}](1-e^{-4y'z}).
\ee

It is important to notice for the following sections that
$K_2(x',y',z',0,0,z)=\left.\diffab{K_2(x',y',z',x,y,z)}{x}\right|_{x=y=0}
=\left.\diffab{K_2(x',y' , z',x,y,z)}{y}\right|_{x=y=0}=0$.

\subsection{Sum rules and large-distance limit}

{}From the results of the previous subsection, we can put the total two-interval
distribution as a sum of 3 terms
\bb\label{E2sum}
\hspace{-1.6truecm}E(x,y,z,t)=E^{(0)}(x,y,z,t)
\\ \nn
\hspace{-1.6truecm}+\int_0^{\infty}\frac{\D x'}{\Lo\sqrt{\pi}}\: E_0(x')K_1(x';x,y,z)
\\ \nn
\hspace{-1.6truecm}+\frac{\sqrt{2\,}}{\Lo^3 {\pi}^{3/2}}\int_{\mathbb{R}_+^3}\!\D x'\D y'\D z'\:
E_0(x',y',z')\wesl{x-x',y-y',z-z'}K_2(x',y',z';x,y,z).
\ee
As a check on our calculations, we consider the particular case of initial conditions given by
a configuration where there is no particle in the system (or $c_0=0$).
In this case, $E(x)$ and $E(x,y,z)$ are always equal to unity for any
time. This means that if we put the conditions $E_0(x')=1$ and $E_0(x',y',z')=1$
into equation \eref{E2sum}  we should recover the result
$E(x,y,z,t)=1$. The first contribution comes from \eref{E2lim} and is
independent of initial conditions. The second contribution comes from
\eref{E21gen} or \eref{E21c0} with $c_0=0$:
\bb\nn
\hspace{-2.0truecm}E(x,y,z,t)=
\erfc\left(\frac{x}{\Lo}\right)\erfc\left(\frac{y}{\Lo}\right)
+\erfc\left(\frac{z}{\Lo}\right)\erfc\left(\frac{x+y+z}{\Lo}\right)
\\ \nn
\hspace{-2.3truecm}-\erfc\left(\frac{x+z}{\Lo}\right)\erfc\left(\frac{y+z}{\Lo}\right)
+\erfc\left(\frac{x}{\Lo}\right)\erf\left(\frac{y}{\Lo}\right)
+\erf\left(\frac{x}{\Lo}\right)\erfc\left(\frac{y}{\Lo}\right)
\\ \nn
\hspace{-2.3truecm}+\erfc\left(\frac{z}{\Lo}\right)\erf\left(\frac{x+y+z}{\Lo}\right)
+\erf\left(\frac{z}{\Lo}\right)\erfc\left(\frac{x+y+z}{\Lo}\right)
\\ \nn
\hspace{-2.3truecm}-\erfc\left(\frac{x+z}{\Lo}\right)\erf\left(\frac{y+z}{\Lo}\right)
-\erf\left(\frac{x+z}{\Lo}\right)\erfc\left(\frac{y+z}{\Lo}\right) \\ \nn
\hspace{-2.3truecm}
+\int_{\mathbb{R}_+^3}\!\!\frac{\sqrt{2\,}\D x'\D y'\D z'}{\Lo^3\sqrt{\pi\,}^3}\:
\wesl{x-x',y-y',z-z'}K_2(x',y',z';x,y,z).
\ee
The terms can be rearranged and we obtain the following equality, since
$E(x,y,z,t)=1$ for all times,
\bb \label{E2coinf}
\hspace{-1.9truecm}
\lefteqn{
\frac{\sqrt{2\,}}{\Lo^3{\pi}^{3/2}}\int_{\mathbb{R}_+^3}\!\!\D x'\D y'\D z'\:
\wesl{x-x',y-y',z-z'}K_2(x',y',z';x,y,z)
}
\\ \nn
\hspace{-1.8truecm}=\erf\left(\frac{x}{\Lo}\right)\erf\left(\frac{y}{\Lo}\right)+\erf\left(\frac{
z }{\Lo}\right)\erf\left(\frac{x+y+z}{\Lo}\right)
-\erf\left(\frac{x+z}{\Lo}\right)\erf\left(\frac{y+z}{\Lo}
\right).
\ee
The last expression gives an identity for the complicated integral involving only
the different weights when the initial functions are set to unity. When
compared to expression \eref{E2lim}, we see that this is the same expression except
for the $\erfc$ functions are replaced by $\erf$. We also
see that all correlators vanish on the empty-lattice, as expected.
{}From the general expression \eref{E2sum} we can compute the limit when
the distance $z$ is large. As seen previously with \eref{E2lim} when the system
is entirely filled with particles, the quantity $E^{(0)}(x,y,z,t)$ is easily
factorised as  $\lim_{z\gg
1}E^{(0)}(x,y,z,t)=\erfc(x/\Lo)\erfc(y/\Lo)=E(x,t)E(y,t)$. In the general case,
for a finite concentration at initial time for example, we can show that the limit is
still valid from \eref{E2sum}. Indeed the two other parts $E^{(1)}$
and $E^{(2)}$ have simpler behavior in this limit. For $E^{(1)}$, the kernel
$K_1$ given in \eref{E21gen} has two dominant terms
\bb\hspace{-1.9truecm}\nn
K_{1,\Lo}(x';x,y,z)\simeq
\erfc(x)
\Big [
e^{-(x'-y)^2}-e^{-(x'+y)^2}
\Big]
+
\erfc(y)
\Big [
e^{-(x'-x)^2}-e^{-(x'+x)^2}
\Big],
\ee
then $E^{(1)}$ can be approximated by
\bb\nn\hspace{-1.9truecm}\label{E21part}
E^{(1)}(x,y,z,t)\simeq
\erfc(x/\Lo)\int_0^{\infty}\!\frac{\D x'}{\sqrt{\pi\,}\,\Lo}\:
E(x',0)
\Big [
e^{-(x'-y)^2/\Lo^2}-e^{-(x'+y)^2/\Lo^2}\Big ]
\\
+\erfc(y/\Lo)\int_0^{\infty}\!\frac{\D x'}{\sqrt{\pi\,}\,\Lo}\:
E(x',0)
\Big [
e^{-(x'-x)^2/\Lo^2}-e^{-(x'+x)^2/\Lo^2}\Big ].
\ee
The other kernel $K_2$ behaves like
\bb\nn
K_{2,\Lo}(x',y',z';x,y,z)\simeq
\Big (1-e^{-4xx'} \Big )\Big (1-e^{-4yy'} \Big ).
\ee
Then
\bb\hspace{-1truecm}\nn
E^{(2)}(x,y,z,t)\simeq
\int_{\mathbb{R}_+^3}\frac{\sqrt{2}\D x'\D y'\D
z'}{\Lo^3\sqrt{\pi}^3}\:
E_0(x',y',z')\we{x-x',y-y',z-z'} \\ \nn
\times \Big (1-e^{-4xx'} \Big )\Big
(1-e^{-4yy'} \Big )
\ee
In the last expression, we set $z''=z'-z$, so that
$E_0(x',y',z')=E_0(x',y',z''+z)$. We also assume that $E_0(x',y',z''+z)
\simeq E_0(x')E_0(y')$, and that the integral over $z''$ can be extended
over the real axis since the lower bound $-z$ is large and negative. Then it
remains an integral over the weight $\we{x-x',y-y',-z''}$, which is independent
of $z$, and which can be performed exactly. We find in particular that
\bb\nn
\int_{-\infty}^{\infty}\frac{\sqrt{2}\D z''}{\Lo\sqrt{\pi}}\:
\we{x-x',y-y',-z''}=e^{-(x-x')^2/\Lo^2-(y-y')^2/\Lo^2}.
\ee
%
\noindent If we multiply this exponential with the kernel $K_2$, we obtain that
\bb\hspace{-1.9truecm}\label{E22part}
E^{(2)}(x,y,z,t)\simeq
\int_0^{\infty}\!\frac{\D x'}{\sqrt{\pi\,}\,\Lo}\:
E(x',0)
\Big [
e^{-(x'-x)^2/\Lo^2}-e^{-(x'+x)^2/\Lo^2}\Big ]
\\ \nn
\times
\int_0^{\infty}\!\frac{\D y'}{\sqrt{\pi\,}\,\Lo}\:
E(y',0)
\Big [
e^{-(y'-y)^2/\Lo^2}-e^{-(y'+y)^2/\Lo^2}\Big ].
\ee

The sum of the expressions \eref{E21part}, \eref{E22part}, combined with the
limit of \eref{E2lim}, gives exactly, after factorization, the expected limit,
which is the generalization of the result seen with \eref{E2lim} in the
particular case of an initially filled system
\bb
\lim_{z\gg 1}E(x,y,z,t)=E(x,t)E(y,t)
\ee
This limit is exact for any given initial condition at any time.

Summarising the contents of this section, we have the decompositon (\ref{gl:E2deco}) of the two-interval probability, 
where the individual terms are given by eqs.~(\ref{E2lim},\ref{E21gen},\ref{K2}), respectively. 

\section{Two-point correlation function}

\subsection{Definition}
The two-point connected correlation function can be defined in the discrete case
as the probability to have two particles separated by the distance $d$
\bb
C_d(t) :=\pr(\bullet\; d\; \bullet) - \pr(\bullet)\pr(\bullet).
\ee
We can express the previous function in terms of the discrete two-interval
functions. Indeed
\bb\nn
\pr(\bullet\; d\; \bullet)+\pr(\circ\; d\;
\bullet)=\pr(\bullet)=1-\pr(\circ)=1-E_{0,1}(d).
\ee
Since $\pr(\circ\; d\; \bullet)+\pr(\circ\; d\; \circ)=\pr(\circ)=E_{1,0}(d)$, we have
$\pr(\circ\; d\; \bullet)=E_{1,0}(d)-E_{1,1}(d)$ and the correlator becomes
\bb\nn
C_d(t) = 1-E_{0,1}(d)-E_{1,0}(d)+E_{1,1}(d) -(1-E_{1,0}(d))^2.
\ee
Using the fact that $E_{0,0}(d)=1$, we can expand $C_d(t)$ up to the second
order in the lattice step $a$, setting $z=da$, and $C(z,t)=C_{d}(t)/a^2$,  so that
\bb
C(z,t)=\left.\partial^2_{xy}E(x,y,z)\right|_{x=0,y=0} -
\left.\partial_xE(x)\right|_{x=0}\left.\partial_yE(y)\right|_{y=0}.
\ee
This is the general expression of the correlation functions, which depends
on the one and two-interval probability functions.

\subsection{Decomposition : general formalism}
There are three contributions to $C(z,t)$ which come  from the second
derivatives with respect to $x$ and $y$ of $E^{(0)}$, $E^{(1)}$ and $E^{(2)}$
respectively. From the two-interval probability eq.~(\ref{gl:E2deco}), we have the decompostion 
\bb \label{eq:Cdeco}
C(z,t)=C^{(0)}(z,t)+C^{(1)}(z,t)+C^{(2)}(z,t),
\ee 
where the different contributions are 
\bb
\hspace{-1cm}C^{(0)}(z,t)&=& \partial^2_{xy}E^{(0)}(x,y,z,t)|_{x=0,y=0} - (\partial_x \erfc(x/\Lo)|_{x=0})^2
\\ \nn
\hspace{-1cm}C^{(1)}(z,t)&=& \partial^2_{xy}E^{(1)}(x,y,z,t)|_{x=0,y=0} - 2\partial_x I(x)|_{x=0} \, \partial_y \erfc(y/\Lo)|_{y=0}
\\ \nn
\hspace{-1cm}C^{(2)}(z,t)&=& \partial^2_{xy}E^{(2)}(x,y,z,t)|_{x=0,y=0} - (\partial_x
I(x)|_{x=0})^2,
\ee
with
\bb
I(x) = \int_0^{\infty} \frac{\D x'}{\sqrt{\pi}\Lo}
E(x',0)\left(e^{-(x'-x)^2/\Lo^2} - e^{-(x'+x)^2/\Lo^2}\right).
\ee

{}From \eref{E2lim}, we obtain the first non-connected contribution of $C^{(0)}(z,t)$, which does not depend on the initial conditions
\bb\nn
\hspace{-0.9truecm}\left.\partial_{xy}^2{E^{(0)}(x,y,z)}\right|_{x=0,y=0}=
\frac{4e^{-z^2/\Lo^2}}{\pi\Lo^2}
\left [
2\sinh(z^2/\Lo^2)+\sqrt{\pi}\frac{z}{\Lo}\erfc\left(\frac{z}{\Lo}
\right)\right]
\ee
and so
\bb  \label{eq:resC0}
\hspace{-1.5truecm}C^{(0)}(z,t)=\frac{4e^{-z^2/\Lo^2}}{\pi\Lo^2}\left[ -e^{-z^2/\Lo^2} +
\sqrt{\pi}\frac{z}{\Lo}\erfc\left(\frac{z}{\Lo}\right)
\right]=:\frac{1}{\Lo^2}f_0(z/\Lo)
\ee
where the scaling function $f_0$ has the following limit behaviour
\bb \label{eq:resC0asy}
f_0(\mathfrak{z}) \simeq \left\{ \begin{array}{ll}
-\frac{\displaystyle 4}{\displaystyle \pi} & \mbox{\rm ~~;~ for~} \mathfrak{z}\to 0 \\[0.16cm]
\frac{\displaystyle 4}{\displaystyle \pi} e^{-\mathfrak{z}^2} & \mbox{\rm ~~;~ for~} \mathfrak{z}\to \infty 
\end{array} \right. .
\ee
Eq.~(\ref{eq:resC0}) is in exact agreement with the result announced in the literature \cite{Alca94,benA98}. 
The asymptotic forms (\ref{eq:resC0asy}) have also
been obtained several times, either for the coagulation-diffusion process \cite{Maye07,Muna06a,Muna06b} or for the
equivalent \cite{Kreb95,Henk95} pair-annihilation-diffusion process, see \cite{Alca94,Gryn95,Mobi02,sch} 
and references therein. 
However, the present discussion is not restricted to the rather special case of an initially fully occupied lattice.
As we shall see, the second and third contributions in (\ref{eq:Cdeco}) 
are corrections to the leading behaviour in the long time limit
(see appendix~F for the details of the proof). 
We obtain a hierarchy in the inverse powers $1/\Lo$, where $\bigl|C^{(0)}(z,t)\bigr|\gg
\bigl|C^{(1)}(z,t)\bigr|\gg \bigl|C^{(2)}(z,t)\bigr|$, with 
a dominant contribution of order $1/\Lo^2$, $1/\Lo^4$ and $1/\Lo^6$,
respectively.

\subsection{Application}

In what follows, we consider the special case of initial conditions $E_0(x')=\exp(-c_0x')$
and $E_0(x',y',z')=\exp(-c_0(x'+y'))$, where $c_0$ is the concentration of
uncorrelated particles at time $t=0$. We assume that initially the two-interval distribution
is independent of the distance between the intervals. It satisfies nevertheless
the condition $E_0(x',y',z'=0)=E_0(x'+y')$. The solutions for the
correlation function given in appendix by \eref{C1gen} and \eref{C2gen}
can be computed except for the triple integral, where the successive gaussian
integrals cannot be expressed, at least to our knowledge, in terms of
known special functions.
Since we are merely interested in the long-time limit and in the influence of the initial
conditions in this limit, we can nevertheless perform an expansion in
$x'/\Lo$ and $y'/\Lo$ inside the kernel $K_{2,\Lo}$ derivatives and the weight function
$\we{-x',-y',z-z'}$ in \eref{C2gen}. This is so because only small values of
$x'$ and $y'$ are relevant
when $c_0\Lo$ is large when these integrands are combined with the exponential
factor $\exp(-c_0(x'+y'))$. The latter function renders the integral finite
after expansion of the weight and kernel.
Indeed the natural small parameter of the expansion is $1/(c_0\Lo)$,
relatively to the other dimensionless parameter $z/\Lo$, and the
series expansion is assumed to break down only in the limit of low
concentration or short times. But the previous result \eref{C2gen} provides a general
form suitable for series expansion of the correlated function with generic initial distributions.
 The other connected part of the correlation function $C^{(1)}(z,t)$, which is defined
in equation \eref{C1gen}, can be computed exactly with the chosen exponential initial condition
$E_0(x')=\exp(-c_0x')$,
or by deriving twice with respect to $x$ and $y$ the previous result given in \eref{E21c0}.
In the asymptotic limit anfn for $c_0\Lo$ large, we obtain
\bb\nn
\hspace{-2.0truecm}
\lefteqn{
C^{(1)}(z,t)=
\frac{8\,z/\Lo}{\sqrt{\pi}\Lo^2(c_0\Lo)^2}\Bigl(-3+2\frac{z^2}{\Lo^2}\Bigr)
\erf(z/\Lo)e^{-z^2/\Lo^2}
} \nn \\
\hspace{-2.4truecm}&+&\frac{16}{\pi \Lo^2(c_0\Lo)^2}\Big
(1-\frac{z^2}{\Lo^2}\Big )e^{-2z^2/\Lo^2}
\nn
+\frac{16z/\Lo}{\sqrt{\pi}\Lo^2(c_0\Lo)^4}
\Big (
15-20\frac{z^2}{\Lo^2}+4\frac{z^4}{\Lo^4}
\Big )
\erf(z/\Lo)e^{-z^2/\Lo^2}
\\
\hspace{-2.4truecm}&+&\frac{32}{\pi\Lo^2(c_0\Lo)^4}
\Big (
-3+9\frac{z^2}{\Lo^2}-2\frac{z^4}{\Lo^4}
\Big )
e^{-2z^2/\Lo^2}
+o\bigl(1/(c_0\Lo)^4\bigr).
\ee
We notice that these first terms contribute to the correlation function
in the large time limit at least like $1/\Lo^4$ times a scaling function of
$z/\Lo$. This is a correction to $C^{(0)}(z,t)$ which behaves like
$1/\Lo^2$ instead.

The last term $C^{(2)}(z,t)$ can also be expanded as a power series of $1/(c_0\Lo)$
whose coefficients are scaling functions of $z/\Lo$
\bb\label{C22}
C^{(2)}(z,t)&=& \frac{16}{\pi\Lo^2}
\exp(-2z^2/\Lo^2)\sum_{k=2}^{\infty}\frac{P_{2k}(z/\Lo)}{(c_0\Lo)^{2k}}
\ee
where the first few polynomials $P_{2k}(z)$ read 
\bb\nn
& &P_{4}(z)=-1-2z^2 \;\; , \;\; P_{6}(z)=12+24z^2-16z^4,
\\ \nn
& &P_{8}(z)=-156-312z^2+528z^4-96z^6,
\\
& &P_{10}(z)=2400+4800z^2-15360z^4+5888z^6-512z^8.
\ee

In general, $C(z,t)$ can be expanded as a series in $1/(c_0\Lo)$,
with coefficients being scaling functions of $z/\Lo$.
Only the contribution $C^{(0)}(z,t)$ is dominant at zero$^{th}$ order in
$1/(c_0\Lo)$, while $C^{(1)}(z,t)$ contributes at order $1/(c_0\Lo)^2$.
Otherwise the dominant part of  $C^{(2)}(z,t)$, as seen in equation \eref{C22}
is of order $1/(c_0\Lo)^4$, therefore smaller than the previous ones. The
correlations are negative over the whole range of considered values of $z$.

\begin{figure}
\centering\resizebox{0.8\columnwidth}{!}
{\includegraphics*{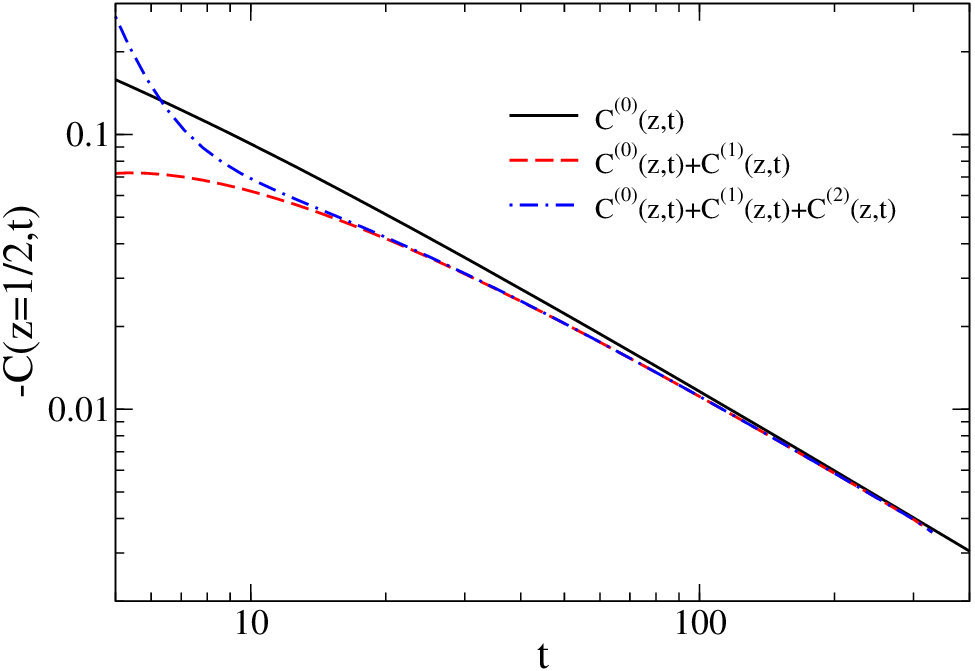}}
\caption{\label{fig3}Time evolution of the connected two-point correlation
function $C(z,t)$ for $c_0=1$ and $z=1/2$. The full solid line shows the leading contribution
$C^{(0)}(1/2,t)$, the dashed line also includes the effect of the leading correction
$C^{(0)}(1/2,t)+C^{(1)}(1/2,t)$ while the dashed-dotted line includes all
contributions $C^{(0)}(1/2,t)+C^{(1)}(1/2,t)+C^{(2)}(1/2,t)$.
}
\end{figure}

In figure~\ref{fig3}, the time-evolution of the correlator $C(1/2,t)$ is illustrated, for initially
uncorrelated particles with a concentration $c_0=1$ and $8D=1$, so that $\Lo=\sqrt{t}$.
In particular, we plot, respectively, the
leading contribution for $t\to\infty$, $C^{(0)}(1/2,t)$, along with the curve resulting when the
first correction $C^{(1)}(1/2,t)$ is included, as well as when both corrective terms $C^{(1)}(1/2,t)+C^{(2)}(1/2,t)$ 
are added. In agreement with our asymptotic estimates, we clearly see that
even for relatively small times, a clear hierarchy 
$\bigl|C^{(0)}(1/2,t)\bigr|\gg \bigl|C^{(1)}(1/2,t)\bigr|\gg \bigl|C^{(2)}(1/2,t)\bigr|$ emerges.

\subsection{Extensions}
Having found the single- and two-hole probabilities, we can immediately derive further quantities of physical interest. 
For example, the effective reaction rate is controlled by the {\em pair probability} $\pr(\bullet\bullet)$ of finding
two particles on neighbouring sites. Similarly, one may define the {\em triplett probability} 
$\pr(\bullet\bullet\bullet)$.  Simple enumeration leads to
\bb
\hspace{-1.0truecm}\pr(\bullet\bullet) &=& E_0 -2E_1 + E_2\stackrel{a\to 0}{=} 
\left.a^2\frac{\partial^2 E(x,t)}{\partial x^2}\right|_{x=0} \nn \\ \nn
\hspace{-1.0truecm}\pr(\bullet\bullet\bullet) &=& 
 E_0 -3E_1 + 2E_2-E_3+ E_{1,1}(1)
\\
&\stackrel{a\to 0}{=}&a^3\left [\left .\partial_{xyz}E(x,y,z)\right|_{x,y,z=0}
-\left .\partial_{xxx}E(x,y,z)\right|_{x,y,z=0}\right ]
\ee
While the pair probability can be expressed in terms of the single-hole probability alone,\footnote{The $x$-dependence of
$\partial_x^2 E(x,t)/c(t)$ describes the {\em interparticle distribution function} 
and has been analysed in detail in the past, see \cite{benA90,benA00} and references therein.}
the triplett probability already depends on a two-hole probability as well. 
In the continuum limit, the pair probability reads, for the three examples of 
initial distributions already considered in figure~\ref{fig1}
\bb
\hspace{-2.2truecm}\pr(\bullet\bullet) = \left\{
\begin{array}{ll} \displaystyle 
c_0^2 \exp\left( \frac{c_0 \Lo}{2}\right) \erfc\left( \frac{c_0 \Lo}{2}\right) & 
\mbox{\rm ~;~ if~} E_0(x)=e^{-c_0 x} \\[0.22cm]
\displaystyle 
\frac{4 c_0^2}{\sqrt{\pi\,}\,} \int_0^{\infty} \!\D x\: e^{-x^2}\,
\frac{1}{(1+c_0\Lo x)^{3}} &
\mbox{\rm ~;~ if~} E_0(x)=(1+c_0 x)^{-1} \\[0.22cm]
0 & \mbox{\rm ~;~ if~} E_0(x)=\erfc\left( \frac{\sqrt{\pi\,}}{2} c_0 x\right) 
\end{array} \right.
\ee
where $c_0$ characterises the width in the initial state. In the first two cases, the long-time behaviour is given by
$\pr(\bullet\bullet)\simeq (2/\sqrt{\pi\,}) (c_0/\Lo) \sim t^{-1/2}$ when $c_0$ is kept fixed. Although the long-time 
behaviour is algebraic, the associated amplitude depends explicitly on the initial distribution, 
in contrast to what we had seen in eq.~(\ref{gl:lemma2.2}) for the particle concentration $c(t)$. 

Similarly, from the results derived in this paper further correlators can be directly obtained, for example
\bb\nn
\hspace{-1.0truecm}\pr(\bullet\bullet d \bullet) &=&   E_0 - 3E_1 + E_2
+E_{1,1}(d) + E_{1,1}(d+1) - E_{2,1}(d) 
\\ 
&\stackrel{a\to 0}{=}&a^3\left [\left .\partial_{xyz}E(x,y,z)\right|_{x,y=0}
-\left .\partial_{xxy}E(x,y,z)\right|_{x,y=0}\right ]
\nn \\ \nn
\hspace{-1.0truecm}\pr(\bullet\bullet d \bullet\bullet) &=& \pr(\bullet\bullet d \bullet) -E_1 + E_2 +E_{1,1}(d+1)+E_{1,1}(d+2) \\
\nn & & - E_{2,1}(d+1)-E_{1,2}(d+1)-E_{1,2}(d)+E_{2,2}(d)
\\ \nn
&\stackrel{a\to 0}{=}&a^4\left [
\left .\partial_{xxyy}E(x,y,z)\right|_{x,y=0}-\left .\partial_{xxyz}E(x,y,z)\right|_{x,y=0}\right .
\\
& &-\left .\left .\partial_{xyyz}E(x,y,z)\right|_{x,y=0}+\left .\partial_{xyzz}E(x,y,z)\right|_{x,y=0}
\right ].
\ee
\section{Conclusions}
We have analysed a method for the computation of time-dependent correlators in the $1D$ coagulation-diffusion process. 
The relevant quantities are the probabilities of finding either a single empty interval of a given size or else
two empty intervals of given sizes and at a given distance from each other. 
These probabilities satisfy simple linear differential or difference 
equations, but working out the explicit solution is rendered difficult through boundary conditions, 
whose symmetry properties
are not the same as those of the differential equations. Rather than circumventing this difficulty, we have shown
how, by analytic continuation to {\em negative} sizes and distances, the full solution may be found for an arbitrary
initial distribution which in turn can be characterised in terms of single- and double-interval probabilities. 

Specifically, we have seen the following:
\begin{enumerate}
\item The leading long-time behaviour of the particle density is explicitly confirmed to be {\em independent} 
of the initial 
conditions, as expected from the field-theoretical renormalisation group \cite{Lee94,Balb95} and in agreement
with earlier calculations using a formulation of the empty-interval method where spatial translation-invariance is not 
immediately built in \cite{benA90,benA00}.\footnote{As discussed in section~5, the various asymptotic estimates
of either the diffusion-coagulation process or the equivalent pair-annihilation-diffusion process are fully reproduced.} 
Our results also allow
to compare directly the continuum limit with the result found for the discrete lattice and we can derive systematically
the finite-time corrections to the leading dynamical scaling behaviour. 
\item The leading long-time behaviour eq.~(\ref{eq:resC0}) of the density-density correlator not only agrees with the 
earlier asymptotic 
estimate \cite{benA98} but is furthermore explicitly shown to be independent of the initial conditions as well. 
The corrections to the leading scaling behaviour have also been analysed and the relative importance of the
initial single- or double-interval probabilities can be quantitatively studied. 
\item By considering general initial conditions, we have identified a natural decomposition, see 
eqs.~(\ref{gl:E2deco},\ref{eq:Cdeco}), into terms which develop a
clear hierarchy in the long-time limit. 
\end{enumerate}
This information will become important in a sequel article where we plan to analyse the ageing behaviour in the
coagulation-diffusion process or generalisations thereof, where methods analogous to the ones developped here
should apply and the single-time correlators studied here will serve as initial values for the two-time correlators
whose dynamical scaling behaviour will be searched for. 

\section*{Acknowledgements}
XD acknowledges the support of the Centre National de la Recherche Scientifique (CNRS) through grant no. 101160.

\appsection{A}{Proof of lemma 3.1}
Starting from the discrete case \eref{E2disc}, we consider the time
differential equation where $n_1=0$, which leads to
\bb\label{E2n10}
\diffab{E_{0,n_2 }(d)}{t}=\diffab{E_{n_2}}{t}.
\ee
The first term is evaluated from \eref{E2disc} and is equal, after some
algebra, to
\bb\nn
\diffab{E_{0,n_2 }(d)}{t}=\Gamma \Big [
2E_{n_2-1}-4E_{n_2}+2E_{n_2+1}
\\
+E_{-1,n_2}(d)+E_{-1,n_2}(d+1)-4E_{n_2}+E_{1,n_2}(d)+E_{1,n_2}(d-1)
\Big ]. \nn
\ee
Using \eref{E1disc}, the right-hand side in \eref{E2n10} is equal to
\bb \nn
\diffab{E_{n_2}}{t}=2\Gamma \Big [
E_{n_2-1}-2E_{n_2}+E_{n_2+1}
\Big ].
\ee
By comparing the last two expressions, we obtain the equality between $n_1=-1$
and $n_1=1$
\bb\label{E2n10bis}
E_{-1,n_2}(d)+E_{-1,n_2}(d+1)+E_{1,n_2}(d)+E_{1,n_2}(d-1)=4E_{n_2}. \nn
\ee
We then differentiate this relation with respect to time, and use
\eref{E2disc} for each of the 4 terms on the l.h.s (also \eref{E1disc} for
the evaluation of $\diffab{E_{n_2}}{t}$) in order to obtain a new relation between
index $n_1=-2$ and index $n_1=2$
\bb\nn
E_{-2,n_2}(d)+2E_{-2,n_2}(d+1)+E_{-2,n_2}(d+2)+
\\ \label{E2n11}
E_{2,n_2}(d)+2E_{2,n_2}(d-1)+E_{2,n_2 } (d-2)=8E_{n_2}. \nn
\ee
By recursion, we can extend this result to any positive index $n_1$
\bb\label{E2n1disc}
\sum_{k=0}^{n_1}\binom{k}{n_1}E_{-n_1,n_2}(d+k)
+
\sum_{k=0}^{n_1}\binom{k}{n_1}E_{n_1,n_2}(d-k)=
2^{n_1+1}E_{n_2}.
\ee
In particular, for $n_2=0$ we recover the result \eref{en}:
$E_{-n_1}+E_{n_1}=2$, by using the relation
$\sum_{k=0}^{n_1}\binom{k}{n_1}=2^{n_1}$.  Equation \eref{E2n1disc}
can be inverted in order to obtain a relation between $E_{-n_1,n_2}(d)$
and functions of positive indices. We use a discrete Fourier transform
\bb\nn
\hspace{-1cm}E_{n_1,n_2}(d)=\int_0^1 \!\D z\; \wit{E}_{n_1,n_2}(z)
e^{ 2\II\pi zd},\;
\wit{E}_{n_1,n_2}(z)=\sum_{d=-\infty}^{\infty}E_{n_1,n_2}(d)
e^{-2\II\pi zd}
\ee
to solve the previous Green's function \eref{E2n1disc}
\bb\hspace{-2cm}
\wit{E}_{-n_1,n_2}(z)\Big (1+e^{2\II\pi z}\Big )^{n_1}=-
\wit{E}_{n_1,n_2}(z)
\Big (
1+e^{-2\II\pi z}
\Big )^{n_1}
+2^{n_1+1}E_{n_2}\sum_{d'}
e^{2\II\pi z(d'-d)} \nn ,
\ee
%
where we used the Dirac relation
\bb\label{Dirac}
\sum_{d'=-\infty}^{\infty}e^{2\II\pi zd'}=\sum_{d'=-\infty}^{\infty}\delta(z- d') .
\ee
Inverting this relation, we obtain
\bb\nn
E_{-n_1,n_2}(d)&=&-\sum_{d'}E_{n_1,n_2}(d')\oint \frac{d\zeta}{2i\pi \zeta}\;\zeta^{d-d'-n_1}
\\
&+&2^{n_1+1}\sum_{d'}E_{0,n_2}(d')\oint \frac{d\zeta}{2i\pi \zeta}
\frac{\zeta^{-d'}}{(1+\zeta)^{n_1}}.
\ee
%
%
The first integral over the variable $z$ gives simply a
Kronecker function $\delta_{d-d'-n_1,0}$, whereas, for the second one, we
the Dirac sum selects the value $z=d'=0$ or $z=d'=1$ when performing the
integration over $z$. We obtain the simple result
\bb\nn
& &\int_0^1 \!\D z\:
\sum_{d'}\frac{e^{2\II\pi zd'}}{(1+e^{2i\pi z})^{n_1}}
=\int_0^1 \!\D z\: \sum_{d'}\delta(z-
d')\frac{1}{(1+e^{2\II\pi z})^{n_1}}
=2^{-n_1}.
\ee
Therefore, we obtain the symmetry relations between the negative and positive indices
$n_1$ (and $n_2$). Relation \eref{E2n123} is directly connected to the first two of the lemma,
by considering the case where the interval lengths are both negative. \hfill q.e.d.

\appsection{B}{Proof of lemma 3.2}
We consider the case $d=0$ in the discrete equation
\eref{E2disc}. Since $E_{n_1,n_2}(0)=E_{n_1+n_2}$, we have
\bb
\diffab{E_{n_1,n_2}(0)}{t}=\diffab{E_{n_1+n_2}}{t}. \nn
\ee
Using \eref{E2disc} the first term gives
\bb\nn
\hspace{-1cm}\diffab{E_{n_1,n_2}(0)}{t}&=&
\Gamma
\Big [
2E_{n_1+n_2+1}-8E_{n_1+n_2}+E_{n_1+1,n_2}(-1)+E_{n_1,n_2+1} (-1)
\\
& &+
2E_{n_1+n_2-1}
+E_{n_1-1,n_2}(1)+E_{n_1,n_2-1}(1)
\Big ], \nn
\ee
whereas the second term gives
\bb
\hspace{-1cm}\diffab{E_{n_1+n_2}}{t}=
2\Gamma
\left (
E_{n_1+n_2-1}-2E_{n_1+n_2}+E_{n_1+n_2+1} \right ). \nn
\ee
By comparing the last two expressions, we obtain
\bb
\hspace{-1cm}E_{n_1+1,n_2}(-1)+E_{n_1,n_2+1} (-1)
+E_{n_1-1,n_2}(1)+E_{n_1,n_2-1}(1)=4E_{n_1+n_2}. \nn
\ee
By deriving successively the last expression with respect to time, we
obtain a general Green's equation for the terms involving a negative and
a positive distance between intervals:
\bb
\hspace{-1cm}\sum_{k=0}^{d}\binom{k}{d}E_{n_1+k,n_2+d-k}(-d)+
\sum_{k=0}^{d}\binom{k}{d}E_{n_1-k,n_2-d+k}(d)
=2^{d+1}E_{n_1+n_2}. \nn
\ee
Notice that the sum of all three indices is always equal to
$n_1+n_2$. As before, we introduce the general Fourier transform
\bb\nn
\hspace{-1cm}E_{n_1,n_2}(d)=\int_0^1 \!\D x\D y\;\wit{E}(x,y,d)
e^{ 2i\pi (n_1x+n_2y)},\;
\\
\wit{E}(x,y,d)=\sum_{n_1,n_2=-\infty}^{\infty}E_{n_1,n_2}(d)
e^{-2i\pi (n_1x+n_2y)} \nn
\ee
to obtain the relation
\bb
\hspace{-1cm}
\wit{E}(x,y,-d)+\wit{E}(x,y,d)
\Big (\frac{e^{-2\II\pi x}+e^{-2\II\pi y}}{e^{2\II\pi
x}+e^{2\II\pi y}}
\Big )^d=
\wit{E}(x,y,0)\frac{2^{d+1}}{(e^{2\II\pi
x}+e^{2\II\pi y})^d}. \nn
\ee
The Fourier inverse then reads explicitly
\bb\nn
\hspace{-1cm}\lefteqn{ \hspace{-1.1truecm}E_{n_1,n_2}(-d)=
\sum_{n_1',n_2'}E_{n_1',n_2'}(0)
\int_0^1 \!\D x\D y\:
\frac{2^{d+1} \exp\bigl(2\II\pi x(n_1-n_1')+2\II\pi y(n_2-n_2')\bigr)}{(e^{2\II\pi x}+e^{2\II\pi y})^d} }
\\
\hspace{-1cm}&-&\sum_{n_1',n_2'}E_{n_1',n_2'}(d)
\int_0^1 \!\D x\D y\:
\Big (
\frac{e^{-2\II\pi x}+e^{-2\II\pi y}}{e^{2\II\pi x}+e^{2\II\pi y}}
\Big )^d
e^{2\II\pi x(n_1-n_1')+2\II\pi y(n_2-n_2')}. \nn
\ee
The second integral over $x$ and $y$ gives simply a double Kronecker function
$\delta_{n_1-n_1'-d,0}\delta_{n_2-n_2'-d,0}$, whereas the first integral can be
transformed as follows
\bb\nn
\hspace{-1.2cm}
\sum_{n_1',n_2'}E_{n_1',n_2'}(0)
\int_0^1 \!\D x\D y\:
\frac{2^{d+1}}{(e^{2\II\pi x}+e^{2\II\pi y})^d}
e^{2\II\pi x(n_1-n_1')+2\II\pi y(n_2-n_2')}
\\ \nn
=
\sum_{n_1',n_2'}E_{n_1'+n_2'}
\int_0^1 \!\D x\D y\:
\frac{2^{d+1}}{(e^{2\II\pi x}+e^{2\II\pi y})^d}
e^{2\II\pi x(n_1-n_1')+2\II\pi y(n_2-n_2')}
\\
=
\sum_{n_1'}E_{n_1'}
\int_0^1 \!\D x\D y\:
\frac{2^{d+1}}{(e^{2\II\pi x}+e^{2\II\pi y})^d}
e^{2\II\pi x(n_1-n_1')+2\II\pi y n_2}
\sum_{n_2'}e^{2\II\pi n_2'(x-y)}. \nn
\ee
The last sum over $n_2'$ can be performed using equality \eref{Dirac}, which
selects $x=y$ over the interval of integration over $y$. Then, the
integration over variable $x$ gives directly a Kronecker function
$\delta_{n_1+n_2-d-n_1',0}$.
Finally we obtain the relation between the index $-d$ and $d$. \hfill q.e.d.

\appsection{C}{Decomposition of the two-interval probability in the three-dimensional space}
We give the detailed calculation for the decomposition of the general solution for the two-hole probability.
Begin by separating the regions with $z'>0$ from those with $z'<0$
\bb\nn
\hspace{-1.9truecm}E_{\ell_0}(x,y,z,t)&=&\sqrt{\frac{2}{\pi^3}\,}
\int_{\mathbb{R}^2}\int_{ 0 }^{ \infty} \!\D x'\D y'\D z' \:
\Bigl[\we{ x-x',y-y',z-z'}
E_{0,\ell_0}(x',y',z')
\\ \hspace{-1.9truecm}&& + \we{x-x',y-y',z+z'}
E_{0,\ell_0}(x',y',-z')\Bigr].
\ee
In the last term, we map  the domain of negative values of $z'$ to the positive values
by using \eref{E2d} in the continuum limit $E_0(x',y',-z')=-E_0(x'-z',y'-z',z')+2E_0(x'+y'-z')$,
where $E_0(x)=E(x,0)$. It is also useful to perform two translations
$x'-z'\rightarrow x'$, $y'-z'\rightarrow y'$ in the first term $E_0(x'-z',y'-z',z')$,
and a translation on the variable $x'\rightarrow  x'-y'+z'$ in the second term $E_0(x'+y'-z')$ so
that
\bb\nn
\hspace{-1.9truecm}E_\Lo(x,y,z)&=&\sqrt{\frac{2}{\pi^3}\,}\int_{\mathbb{R}^2}\int_{ 0 }^{ \infty}\!\D x'\D y'\D z'\:
\we{ x-x',y-y',z-z'}
E_{0,\Lo}(x',y',z')
\\ \nn& &- \we{x-x'-z',y-y'-z',z+z'} E_{0,\Lo}(x',y',z')
\\
& &+ 2\we{ x-x'+y'-z',y-y',z+z'}E_0(x').
\ee
In the third term containing $E_0(x')$, we can now perform the integration
on $y'$ and $z'$, since the function $\we{}$ is a Gaussian kernel.
When $x'$ is negative, we use also $E_0(x')=2-E_0(-x')$ in order to perform the
mapping onto the positive axis.
After rearranging the different terms we obtain
\bb\nn
\hspace{-1.5truecm}E_\Lo(x,y,z)&=&\erfc(z)\erfc(x+y+z)
\\ \nn & &+\erfc(z)\int_0^{\infty}
\frac{\D x'}{\sqrt{\pi}}
\left[
e^{-(x+y+z-x')^2}-e^{-(x+y+z+x')^2}
\right]
E_{0,\Lo}(x')
\\ \nn
\hspace{-1.5truecm}& &+
\sqrt{\frac{2}{\pi^3}\,}\int_{\mathbb{R}^2}\int_{ 0 }^{ \infty}\!\D x'\D y'\D z'\:
\Big [
\we{x-x',y-y',z-z'}
\\ \label{E2va}
\hspace{-1.5truecm}\quad \quad \quad& &-\we{x-x'-z',y-y'-z',z+z'}
\Big ]
E_{0,\Lo}(x',y',z')
\ee
In the last term, the difference between the two weight functions $\we{}$
can be simplified by noticing that
\bb\nn
\we{x-x',y-y',z-z'}-\we{x-x'-z',y-y'-z',z+z'}
\\ \label{relation-z}
=\we{x-x',y-y',z-z'}(1-e^{-4zz'})
:=\wet{x',y',z'}.
\ee
{}From the last multiple integral \eref{E2va}, we can divide the integration
domain over $x'$ and $y'$ into 4 parts: $E_0(x',y',z')$, $E_0(-x',y',z')$,
$E_0(x',-y',z')$, and $E_0(-x',-y',z')$, with $(x',y',z')$ all positive
variables. $E_0(-x',y',z')$ can be transformed using the following steps
and \eref{E2n12} and \eref{E2d}
\bb\nn
\hspace{-1.8truecm}E_0(-x',y',z')&=&2E_0(y')-E_0(x',y',z'-x')
\\ \nn
&=&
2E_0(y')-\theta(z'-x')E_0(x',y',z'-x')
\\
& &-\theta(x'-z')\Big [
2E_0(y'+z')-E_0(z',-x'+y'+z',x'-z')
\Big ]
\ee
then
\bb\nn
\hspace{-1.5truecm}E_0(-x',y',z')&=&2E_0(y')-\theta(x'-z')2E_0(y'+z')
\\ \nn& &+
\theta(x'-z')\theta(x'-y'-z')2E_0(z')
\\ \nn
& &-\theta(z'-x')E_0(x',y',z'-x')
\\ \nn
& &+\theta(x'-z')\theta(-x'+y'+z')E_0(z',-x'+y'+z',x'-z')
\\ \label{partx}
& &-\theta(x'-z')\theta(x'-y'-z')E_0(z',x'-y'-z',y').
\ee
All arguments of the functions appearing in the last expression are
now positive.
An analogous analysis is done for $E_0(x',-y',z')$, which is symmetric by inverting
$x'$ and $y'$
\newpage
\typeout{*** saut de page ***}
\bb\nn
\hspace{-1.5truecm}E_0(x',-y',z')&=&2E_0(x')-\theta(y'-z')2E_0(x'+z')
\\ \nn& &+
\theta(y'-z')\theta(-x'+y'-z')2E_0(z')
\\ \nn
& &-\theta(z'-y')E_0(x',y',z'-y')
\\ \nn
& &+\theta(y'-z')\theta(x'-y'+z')E_0(x'-y'+z',z',y'-z')
\\ \label{party}
& &-\theta(y'-z')\theta(-x'+y'-z')E_0(-x'+y'-z',z',x').
\ee
Finally, using \eref{E2n123} and \eref{E2d} the last expression $E_0(-x',-y',z')$ is
transformed into
\bb\nn
\hspace{-1.9truecm}E_0(-x',-y',z')&=&4-2E_0(x')-2E_0(y')+E_0(x',y',z'-x'-y')
\\ \nn
&=&4-2E_0(x')-2E_0(y')+\theta(z'-x'-y')E_0(x',y',z'-x'-y')
\\ \nn& &+
\theta(x'+y'-z')\Big [
2E_0(z')-E_0(z'-y',z'-x',x'+y'-z')
\Big ].
\ee
In the last term, $z'-y'$ and $z'-x'$ can be either positive or negative, which
leads us to analyse the other possibilities
\bb\label{partxy}
\lefteqn{ \hspace{-2.3truecm}E_0(-x',-y',z')
={4}-2E_0(x')-2E_0(y')+\theta(x'+y'-z')2E_0(z')
}
\\ \nn
\hspace{-2.3truecm}&\hspace{-1.5truecm}+&\hspace{-1.0truecm}\theta(z'-x'-y')E_0(x',y',z'-x'-y') \\ \nn
\hspace{-2.3truecm}&\hspace{-1.5truecm}-&\hspace{-1.0truecm}\theta(x'+y'-z')
\Big [
\theta(z'-y')\theta(z'-x')E_0(z'-y',z'-x',x'+y'-z') \\ \nn
\hspace{-2.3truecm}&\hspace{-1.5truecm}+&\hspace{-1.0truecm}\theta(y'-z')\theta(z'-x')\{-E_0(y'-z',z'-x',x')+2E_0(z'-x') \}
\\ \nn
\hspace{-2.3truecm}&\hspace{-1.5truecm}+&\hspace{-1.0truecm}\theta(z'-y')\theta(x'-z')\{-E_0(z'-y',x'-z',y')+2E_0(z'-y') \}
\\ \nn
\hspace{-2.3truecm}&\hspace{-1.5truecm}+&\hspace{-1.0truecm}
\theta(y'-z')\theta(x'-z')\{{4}-2E_0(y'-z')-2E_0(x'-z')+E_0(y'-z',x'-z',z') \}
\Big ].
\ee
Here, we have (i) terms without any empty-interval probability, 
(ii) terms which refer to the single-interval probabilities only
and (iii) terms which contain also the  two-interval probabilities. The first
kind of terms is needed for an initially fully occupied lattice (see section 4).

\appsection{D}{Simplification of some integrals for the initial one-interval contribution}
We show how to simplify the integrals $I_1,\dots,I_6$ of
section 4.2.
In order to arrive at the previous partial result, it is useful to
note that a function $\theta(x'-z')$ (in the case when it involves an integral
over the one interval distribution $E_0$ which does not depend on $x'$) with
$x'$ and $z'$ positive can be simplified and replaced by unity if we perform the
translation $x''=x'-z'$. The integration over $x''$ extends then from $-z'<0$ to
$\infty$, however the function $\theta$ restricts the integration to the interval $x''>0$.
Therefore the limits of the integration are unchanged under this procedure, and
the function $\theta$ can be set to unity. In the opposite case, where we have a
contribution such as $E_0(x')\theta(x'-z')$ for example, a translation on $x'$
will change the argument of $E_0$, which does not conserve the form \eref{E21a}.
Instead, we write $\theta(x'-z')=1-\theta(z'-x')$ and apply the previous
translation to the positive variable $z'=z''+x'$. This changes the arguments of
the $\wet{\cdots}$ functions only and not $E_0$. Using \eref{relation-z2}, the
different groups of terms $I_1,\dots,I_4$ can be simplified
\bb
\nn
I_1&=&\wet{x',-y',z'}-\wet{x'-z',-y'-z',z'} \\ \nn
&=&\wet{x',-y',z'}+\wet{x',-y',-z'}
\\
I_2&=&\wet{-y',x',z'}-\wet{-y'-z',x'-z',z'} \\ \nn
&=& \wet{-y',x',z'}+\wet{-y',x',-z'}
\\ \nn
I_3&=&\wet{-x'-z',-y'-z',z'}-\wet{-x',-y',z'} \\ \nn
&=&-\wet{-x',-y',-z'}-\wet{-x',-y',z'}
\\ \nn
I_4&=&\wet{-y'-z',-x'-z',z'}-\wet{-y',-x',z'} \\ \nn
&=& -\wet{-y',-x',-z'}-\wet{-y',-x',z'}.
\ee
These relations allow us to extend the integration over $z'$ from $-\infty$
to $+\infty$ since they combine each time two terms depending on $z'$ and
$-z'$. Since
the variable $z'$ appears only in the gaussian weights,
the integration over $z'$ gives new gaussian exponentials. Then integration
over $y'$ gives erf or erfc functions, which are combined with $E_0(x')$.
Terms $I_5$ and $I_6$ can be combined together. The first term of $I_5$ can be
transformed as
\bb\nn
\int_{\mathbb{R}_+^3} \!\D x'\D y'\D z'\; E_0(x')
\theta(-x'-y'+z')\wetsl{y',-z',x'}
\\ \nn
=
\int_{\mathbb{R}_+^3} \!\D x'\D y'\D z'\; E_0(x')
\theta(-x'+y'-z')\wetsl{z',-y',x'}
\\ \nn
=\int_0^{\infty}\!\D x'\: E_0(x')
\int_0^{\infty}\!\D y'\int_{-\infty}^{0}\!\D z'\:
\theta(-x'+y'+z')\wetsl{-z',-y',x'}.
\ee
The second term gives
\bb\nn
\int_{\mathbb{R}_+^3} \!\D x'\D y'\D z'\; E_0(x')
\theta(-x'-y'+z')\wetsl{-z',y',x'}=
\\ \nn
=\int_0^{\infty}\!\D x'\: E_0(x')
\int_{-\infty}^{0}\!\D y'\int_0^{\infty}\!\D z'\:
\theta(-x'+y'+z')\wetsl{-z',-y',x'}.
\ee
Finally, consider $I_6$, which is equal to
\bb\nn
\int_0^{\infty}\!\D x'\: E_0(x')
\int_0^{\infty}\!\D y'\int_0^{\infty}\!\D z'\:
\theta(-x'+y'+z')\wetsl{-z',-y',x'}
\ee
and can be combined with the previous two terms. Noticing that
$\theta(-x'+y'+z')=0$ in the domain where $y'$ and $z'$ are both negative,
we can write the integral over $I_5$ and $I_6$ as follows
\bb\nn
\hspace{-1cm}\sqrt{\frac{8}{\pi^3}\,}
\int_{\mathbb{R}_+^3}\!\D x'\D y'\D z'\;
E_{0,\Lo}(x')
\Big (
I_5+I_6
\Big )
\\ \nn
\hspace{-1cm}=2\sqrt{2}
\int_0^{\infty}\frac{\D x'}{\sqrt{\pi}}\: E_{0,\Lo}(x')
\int\!\!\!\!\int_{-\infty}^{\infty}\frac{\D y'\D z'}{\pi}\:
\theta(-x'+y'+z')\wet{-z',-y',x'}
\\ \nn
\hspace{-1cm}=2\sqrt{2}\int_0^{\infty}\frac{\D x'}{\sqrt{\pi}}\: E_{0,\Lo}(x')
\int\!\!\!\int_{-\infty}^{\infty}\frac{\D y'\D z'}{\pi}\:
\theta(-x'+z')\wet{y'-z',-y',x'}.
\ee
Then, it is possible to perform the integration over $y'$ and $z'$ successively.

\appsection{E}{Simplification of some integrals for the initial two-interval contribution}
In this appendix, we show how to simplify the expression of the contribution 
which depends on the initial two-interval probability.\\
There are therefore 12 terms involving the weights $\wet{\cdots}$.
After performing different translation operations, we arrive at the expression
\bb\nn
E^{(2)}_\Lo(x,y,z)= \sqrt{\frac{2}{\pi^3}\,}
\int_{\mathbb{R}_+^3}
\!\D x'\D y'\D z' \; E_{0,\Lo}(x',y',z')\times
\\ \nn
\Big \{
\wet{x',y',z'}-\wet{-y'-z',-x'-z',z'}
\\ \nn
+\wet{-x',-y',x'+y'+z'}-\wet{-x'-z',-y'-z',x'+y'+z'}
\\ \nn
+\wet{-x'-y'-z',-z',x'+z'}-\wet{-x',y',x'+z'}
\\ \nn
+\wet{-z',-x'-y'-z',y'+z'}-\wet{x',-y',y'+z'}
\\ \nn
+\wet{-x'-z',y'+z',x'}-\wet{-x'-y'-z',z',x'}
\\ \label{E22a}
+\wet{x'+z',-y'-z',y'}-\wet{z',-x'-y'-z',y'}
\Big \}.
\ee
We can pair the terms which appear on each line of the previous relation
noticing that
\bb
\hspace{-1.0cm}\wet{u-w,v-w,w-(u+v)}=\wet{u,v,w-(u+v)}e^{-4w(x+y+z)}.
\ee
This yields the following relations
\bb\nn
\hspace{-1.0cm}\wet{-y'-z',-x'-z',z'}=\wet{x',y',z'}e^{-4(x'+y'+z')(x+y+z)},
\\ \nn
\hspace{-1.0cm}\wet{-x'-z',-y'-z',x'+y'+z'}=\wet{-x',-y',x'+y'+z'}e^{-4z'(x+y+z)},
\\ \nn
\hspace{-1.0cm}\wet{-x'-y'-z',-z',x'+z'}=\wet{-x',y',x'+z'}e^{-4(y'+z')(x+y+z)},
\\ \nn
\hspace{-1.0cm}\wet{-z',-x'-y'-z',y'+z'}=\wet{x',-y',y'+z'}e^{-4(x'+z')(x+y+z)},
\\ \nn
\hspace{-1.0cm}\wet{-x'-y'-z',z',x'}=\wet{-x'-z',y'+z',x'}e^{-4y'(x+y+z)},
\\ \nn
\hspace{-1.0cm}\wet{z',-x'-y'-z',y'}=\wet{x'+z',-y'-z',y'}e^{-4x'(x+y+z)}.
\ee
Moreover, we have
\bb\nn
\hspace{-1.5cm}\wet{x',y',z'}=\we{x-x',y-y',z-z'}(1-e^{-4z'z}),
\\ \nn
\hspace{-1.5cm}\wet{-x',-y',x'+y'+z'}=\we{x-x',y-y',z-z'}e^{-4x'x-4y'y}[1-e^{-4(x'+y'+z')z}],
\\ \nn
\hspace{-1.5cm}\wet{-x',y',x'+z'}=\we{x-x',y-y',z-z'}e^{-4x'x}[1-e^{-4(x'+z')z}],
\\ \nn
\hspace{-1.5cm}\wet{x',-y',y'+z'}=\we{x-x',y-y',z-z'}e^{-4y'y}[1-e^{-4(y'+z')z}],
\\ \nn
\hspace{-1.5cm}\wet{-x'-z',y'+z',x'}=\we{x-x',y-y',z-z'}e^{-4x'x-4z'(x+z)}(1-e^{-4x'z})
\\ \nn
\hspace{-1.5cm}\wet{x'+z',-y'-z',y'}=\we{x-x',y-y',z-z'}e^{-4y'y-4z'(y+z)}(1-e^{-4y'z}).
\ee
%

\appsection{F}{Two-point correlation function}
In this appendix, we show how the terms depending on initial conditions are
corrections to the leading behaviour in the long-time limit.\\
We first consider the contribution $C^{(1)}(z,t)$ which can be evaluated by
deriving the kernel in \eref{E21gen}
\bb\nn
\hspace{-2.0truecm}C^{(1)}(z,t)&=&\int_0^{\infty}\frac{\D x'}{\sqrt{\pi}\Lo^3}\;E(x',0)\left(
\erf(z/\Lo)\Big \{2-4(x'+z)^2/\Lo^2\Big \}e^{-(x'+z)^2/\Lo^2} \right.
\\ \label{C1gen}
\hspace{-2.0truecm}& &-\erf(z/\Lo) \Big \{2-4(x'-z)^2/\Lo^2\Big \} e^{-(x'-z)^2/\Lo^2}
\\ \nn
\hspace{-2.0truecm}& &\left. +\frac{4}{\sqrt{\pi}\Lo}\Big \{
(2x'-z)e^{-(x'-z)^2/\Lo^2}+(2x'+z)e^{-(x'+z)^2/\Lo^2}
\Big \}e^{-z^2/\Lo^2}\right)\\ \nn
\hspace{-2.0truecm}&=:& \int_0^{\infty}\frac{\D x'}{\sqrt{\pi}\Lo^3}\;E(x',0) L_1(x',z)
\ee
In order to obtain the term $C^{(2)}(z,t)$, we use the important property
previously seen that
$K_2(x',y',z',0,0,z)=\left.\diffab{K_2(x',y',z',x,y,z)}{x}\right|_{x=y=0}
=\left.\diffab{K_2(x',y' , z',x,y,z)}{y}\right|_{x=y=0}=0$, which can be
evaluated from \eref{K2}, so that
\bb\nn
\hspace{-2.4truecm}
C^{(2)}(z,t)&=& \frac{\sqrt{2\,}}{\pi^{3/2} \Lo^3} \int_{\mathbb{R}_+^3} \!\D x'\,\D y'\,\D z'\: 
E_{0}(x',y',z')\wesl{-x',-y',z-z'} \times
\\ \nn
\hspace{-2.4truecm}& &\quad \left\{ 16e^{-4z'z/\Lo^2}
\Big [
x'y'
\Big (
e^{4z'z/\Lo^2}+e^{-4(x'+y'+z')z/\Lo^2} \Big ) \right.
\\ \nn
\hspace{-2.4truecm}& & \quad \; -(x'+z')(y'+z')
\Big ( 1+e^{-4(x'+y')z/\Lo^2} \Big )
\Big ]
\\ \nn
\hspace{-2.4truecm}& & \quad \; \left. +16e^{-4z'z/\Lo^2}z'(x'+y'+z')\Big ( e^{-4x'z/\Lo^2}+e^{-4y'z/\Lo^2} \Big ) \right\}/\Lo^4
\\ \label{C2gen}
\hspace{-2.4truecm}& &- \left(\int_ 0^{\infty} \frac{\D x}{\sqrt{\pi}\Lo^3} \; 4x\,E(x,0) \; e^{-x^2/\Lo^2}\right)^2
\\ \nn
\hspace{-2.4truecm}&=:& \frac{\sqrt{2\,}}{\pi^{3/2} \Lo^3} \int_{\mathbb{R}_+^3}
\!\D x'\,\D y'\,\D z'\: E_{0}(x',y',z')\wesl{-x',-y',z-z'
} L_2(x',y',z',z) 
\\ \nn
\hspace{-2.4truecm}& & - \left(\int_ 0^{\infty} \!\D x\:\frac{4x}{\sqrt{\pi}\Lo^3} \; E(x,0) \; e^{-x^2/\Lo^2}\right)^2
\ee
Therefore, we can sum up the different contributions to express the exact
correlation function. In order to study the effects of
the initial conditions in the long-time behaviour on the scaling form, we can
isolate the dominant contribution for each of the three terms $C^{(0)}$,
$C^{(1)}$, $C^{(2)}$, depending on the properties of the initial distribution
function for the two intervals and in the case where $z/\Lo$ is small. For
$C^{(0)}$, we obtain easily
\bb\label{C0}
C^{(0)}(z,t) &=& \frac{1}{\Lo^2} f_0(z/\Lo),\;{\rm with}\;
\;\;f_0(z/\Lo)
\stackrel{\Lo \gg 1}{\simeq} -\frac{4}{\pi}
\ee

For $C^{(1)}$, when $\Lo$ is large, we can perform a Taylor expansion of
$L_1(x',z)$ around $x' =0$, and since $L_1(0,z)=0$ we obtain
\bb\label{C1}
C^{(1)}(z,t) \simeq \frac{f_1(z/\Lo)}{\Lo^4}\int_0^{\infty}\!\D x\;E(x,0)\, x
\ee
where the scaling function $f_1$ is given by the expression
\bb\hspace{-1cm}\nn
f_1(\frac{z}{\Lo})=\frac{8}{\pi} e^{-z^2/\Lo^2}\Big [
2\left(1-\frac{z^2}{\Lo^2}\right)e^{-z^2/\Lo^2}+\sqrt{\pi}\frac{z}{\Lo}
\left(2\frac{z^2}{\Lo^2}-3\right)\erf\left(\frac{z}{\Lo}\right)
\Big ]
\\
\stackrel{\Lo \gg z}{\simeq}
 \frac{16}{\pi}\Big (1-6\frac{z^2}{\Lo^2}\Big )
\ee
The quantity $E_1:=\int_0^{\infty}\!\D x\:E(x,0)x$ is related to the second
moment of
the interval distribution $P(x,0)$ \eref{EP} after performing an integration by
parts
\bb\label{Moy1}
E_1=\int_0^{\infty}\!\D x\;E(x,0)\,x=\frac{1}{2}\int_0^{\infty}\!\D x\;P(x,0)\, x^2.
\ee

Concerning the dominant behaviour of $C^{(2)}$, we notice that the
gaussian weight $\wesl{-x',-y',z-z'}$ is peaked around the value
$(x',y',z')=(0,0,z)$. We can therefore begin to perform a Taylor expansion of
$E_{0}(x',y',z')$ around the $z'=z$ and integrate the variable $z'$ on
the real axis if $z$ is far enough from the value zero, which also is satisfied
if $z/\Lo$ small.
\bb\hspace{-2cm}\nn
\lefteqn{C^{(2)}(z,t) \simeq \int_{\mathbb{R}_+^2} \D x'\,\D
y'E_{0}(x',y',z)\underbrace{\int_{\mathbb{R}}\frac{\sqrt{2}\D
z'}{(\sqrt{\pi})^3\Lo^5} \wesl{-x',-y',z-z'} L_2(x',y',z',z)}_{M^{}_0(x',y',z)}
}
\\ \nn \hspace{-2cm}
&+& \int_{\mathbb{R}_+^2} \D x'\,\D y'
\partial_zE_{0}(x',y',z)\underbrace{\int_{\mathbb{R}}\frac{\sqrt{2}\D
z'}{(\sqrt{\pi})^3\Lo^5}(z'-z)\wesl{-x',-y',z-z'}
L_2(x',y',z',z)}_{M^{}_1(x',y' ,z)}
\\ \label{C2} \hspace{-2cm}
& &- \left(\int_ 0^{\infty} \!\D x\: \frac{4x}{\sqrt{\pi}\Lo^3} \; E(x,0) \; e^{-x^2/\Lo^2}\right)^2
\ee
For $\Lo$ large, the function $M_0$ can be expanded as a series in $x'$ and $y'$
to obtain the dominant term
\bb\nn
M_0(x',y',z)\simeq \frac{x'y'}{\Lo^6}f_2^{(0)}(\frac{z}{\Lo}) \;\; , \;\; 
f_2^{(0)}(u)=-\frac{32}{\pi}e^{-2 u^2}\Big (
1+2 u^2-e^{2 u^2} \Big ).
\ee
The next function $M_1$ in the development can be expanded as
\bb\nn
M_1(x',y',z)\simeq \frac{x'y'}{\Lo^6}f_2^{(1)}(\frac{z}{\Lo}) \;\; , \;\; 
f_2^{(1)}(u)=-u f_2^{(0)}(u).
\ee
The last term in \eref{C2} can be approximated by
\bb \nn
\left(\int_ 0^{\infty} \!\D x\:\frac{4 x}{\sqrt{\pi}\Lo^3} \; E(x,0) \;
e^{-x^2/\Lo^2}\right)^2 \simeq \frac{16}{\pi \Lo^6}
E_1^2.
\ee
where $E_1$ is the average quantity of the single interval distribution, given
in \eref{Moy1}. 

Hence, $C^{(2)}$ can be expanded as a series of the inverse diffusion
length involving integrals of the two-interval distribution and
related moments, as long as these integrals are not diverging:
\bb \nn
\hspace{-2cm}
C^{(2)}(z,t)\simeq
\frac{f_2^{(0)}(z/\Lo)}{\Lo^6}\int_{\mathbb{R}_+^2} \D x'\D y'\,
x'\,y'\;E_{0}(x',y',z) \\ \label{C2-2}
+
\frac{f_2^{(1)}(z/\Lo)}{\Lo^6}\int_{\mathbb{R}_+^2} \D x'\D y'\,
x'\,y'\;\partial_z E_{0}(x',y',z)
- \frac{16}{\pi \Lo^6}E_1^2.
\ee
The expansion \eref{C2-2} has the advantage that if the initial condition 
$E_{0}(x,y,z)$ does not depend on the distance $z$ between the two intervals
$x$ and $y$, then the second term of the development involving 
$\partial_z E_{0}(x,y,z)$ vanishes.
However, from the previous results \eref{C0}, \eref{C1} and \eref{C2-2}, we obtain
a hierarchy in the inverse powers $1/\Lo$, where $\bigl|C^{(0)}(z,t)\bigr|\gg
\bigl|C^{(1)}(z,t)\bigr|\gg \bigl|C^{(2)}(z,t)\bigr|$, with 
a dominant contribution of order $1/\Lo^2$, $1/\Lo^4$ and $1/\Lo^6$, successively. {}From the previous asymptotic results,
it is also easy to check  that in every case, the sign of the correlation contributions are respectively 
$C^{(0)}<0$, $C^{(1)}>0$ and $C^{(2)}<0$.
In particular, if $z\ll \Lo$, the dominant contribution to
$C_2$ comes from the last term of \eref{C2-2}, since the scaling functions
$f_2^{(0)}$ and $f_2^{(1)}$ are rapidly decreasing functions of $z/\Lo$. Hence
\bb
C^{(2)}(z,t) \stackrel{\Lo \gg 1}{\simeq} -\frac{16}{\pi \Lo^6}E_1^2 .
\ee

\appsection{G}{Some integral identities}
We list some identities involving the kernel ${\cal W}$ and the error function $\erf$, which are used in the main text.
\bb
\hspace{-1cm}\sqrt{2}
\int\!\!\!\int\!\!\!\int_{-\infty}^{\infty}\frac{\D x'\D y'\D z'} { \sqrt{\pi}^3 }\:
\wet{x',y',z'}=0
\ee
\bb
\hspace{-1cm}\sqrt{2}\int_0^{\infty}\frac{\D x'}{\sqrt{\pi}}
\int\!\!\!\int_{-\infty}^{\infty}\frac{\D y'\D z'} { \pi }\:
\wet{x',y',z'}
=\frac{1}{2}\Big [
\erf(x)-\erf(x+z) \Big ]
\\
\hspace{-1cm}\sqrt{2}\int_0^{\infty}\frac{\D y'}{\sqrt{\pi}}
\int_{-\infty}^{\infty}\frac{\D x'\D z'} { \pi }\:
\wet{x',y',z'}
=\frac{1}{2}\Big [
\erf(y)-\erf(y+z) \Big ]
\ee
\bb
\hspace{-1cm}\sqrt{2}\int\!\!\!\int_{-\infty}^{\infty}\frac{\D x'\D y'}{\pi}
\int_{0}^{\infty}\frac{\D z'} { \sqrt{\pi\,} }\:
\wet{x',y',z'}=\erf(z)
\ee
\bb\nn
\hspace{-1cm}4\sqrt{2}\int\!\!\!\int_{0}^{\infty}\frac{\D x'\D y'}{\pi}
\int_{-\infty}^{\infty}\frac{\D z'} { \sqrt{\pi\,} }\:
\wet{x',y',z'}=\erf(x)-\erf(x+z)
\\
+\erf(y)-\erf(y+z)+
\erf(x)\erf(y)-\erf(x+z)\erf(y+z)
\ee
\bb\nn
\hspace{-1cm}4\sqrt{2}
\int_{-\infty}^{\infty}\frac{\D y'}{\sqrt{\pi\,}}
\int\!\!\!\int_{0}^{\infty}\frac{\D x' \D z'}{\pi}\:
\we{x+x'+y'-z',y-y',z+z'}
\\ =\erfc(z)\erfc(x+y+z)
\ee
\bb\nn
\hspace{-1cm}2\sqrt{2}
\int_{-\infty}^{\infty}\frac{\D y'}{\sqrt{\pi\,}}
\int\!\!\!\int_{0}^{\infty}\frac{\D x'\D z'}{\pi}\:
\wet{-x'-y'+z',y',-z'}
\\ =-\erf(z)\erfc(x+y+z)
\ee

\newpage

\section*{References}

\end{document}